\documentclass[aps,12pt,showpacs,showkeys]{iopart}

\usepackage[english]{babel}
\usepackage{amstext}
\usepackage{graphicx}
\graphicspath{{figures/}}
\usepackage{simplewick}
\usepackage{verbatim}

\newcommand{\be}{\begin{eqnarray}}
\newcommand{\ee}{\end{eqnarray}}
\newcommand{\bdm}{\begin{displaymath}}
\newcommand{\edm}{\end{displaymath}}
\newcommand{\ds}{\displaystyle}
\newcommand{\ba}{\begin{array}}
\newcommand{\ea}{\end{array}}
\newcommand{\pa}[1]{\left(#1\right)}
\newcommand{\paq}[1]{\left[#1\right]}
\newcommand{\pag}[1]{\left\{#1\right\}}

\newcommand{\dpa}{\partial}
\newcommand{\K}{{\bf k}}

\newcommand{\X}{{\bf x}}

\newcommand{\az}{\mathcal{S}}
\newcommand{\dddot}[1]{\stackrel{\ldots}{#1}}
\newcommand{\0}{{\bf 0}}
\newcommand{\1}{{\bf 1}}
\newcommand{\2}{{\bf 2}}

\begin{document}
\title{Effective field theory methods to model compact binaries}

\author{Stefano Foffa$^{\rm 1}$ and Riccardo Sturani$^{\rm 2}$}

\address{$(1)$ D\'epartement de Physique Th\'eorique, Universit\'e de 
  Gen\`eve, CH-1211 Geneva, Switzerland\\
  $(2)$ ICTP South American Institute for Fundamental Research\\
  Instituto de F\'\i sica Te\'orica, Universidade Estadual Paulista,
  Sao Paulo, SP 011040-070, Brasil}

\ead{stefano.foffa@unige.ch, sturani@ift.unesp.br}

\begin{abstract}
In this short review we present a self-contained exposition of the effective 
field theory method approach to model the dynamics of gravitationally bound 
compact binary systems within the post-Newtonian approximation to General
Relativity. 
Applications of this approach to the conservative sector,
as well as to the radiation emission by the binary system are discussed in 
their salient features. Most important results are discussed in a pedagogical 
way, as in-depths and details can be found in the referenced papers.
\end{abstract}


\maketitle

\section{Introduction}
\label{se:intro}
The existence of gravitational waves (GW) is an unavoidable prediction of 
General Relativity (GR) and astrophysical objects bound in binary systems are 
prototypical, even though not exclusive, sources of GWs.
The precise evidence of a system emitting GWs comes from the celebrated 
``Hulse-Taylor'' binary pulsar \cite{Hulse:1974eb}, whose orbital decay rate
is in agreement with the GR prediction to about one part in a thousand 
\cite{Weisberg:1984zz}, see also \cite{Burgay:2003jj,Kramer:2009zza,Wolszczan:1991kj,Stairs:2002cw}
for more examples of observed GW emission from pulsar binary systems.

A network of earth-based, kilometer-sized GW observatories
is currently under development with the goal of detecting GWs: the two Laser 
Interferomenter
Gravitational-Wave Observatories (LIGO) in the US and the French-Italian Virgo 
interferometer in Italy have been 
taking data at unprecedented sensitivities for several years, see e.g.~\cite{Aasi:2012rja}
for recent results, and are now 
undergoing upgrades to their advanced stage, see e.g.~\cite{Sturani:2012li} for
a recent review
(another smaller detector belonging to the network is the 
German-British Gravitational Wave Detector GEO600). 
The gravitational detector network is planned to be joined by the
Japanese detector KAGRA by the end of this decade \cite{Aso:2013eba}
and by an additional
interferometer in India by the beginning of the next decade.
The advanced detector era is planned to start in the year 2015 and it is 
expected that few years will be necessary to reach planned sensitivity, which 
should allow several detections of GW events per year \cite{Abadie:2010cf}.

Compact binary systems offer a privileged setting where to confront GR with 
observations, and their dynamics has been the object of 
intensive studies since the advent of GR. Here we focus on the 
\emph{post-Newtonian} (PN)
approximation to GR, see e.g.~\cite{lrr-2006-4} for a review, which consists in
a perturbative expansion around Minkowki space.
The expansion parameter is the relative velocity $v$ of the binary constituents
\footnote{We posit the speed of light $c=1$.},
or equivalently the gravitational field strentgh $G_N M/r$ (where $G_N$ is the 
standard Newton constant, $M$ the total mass of the binary system, and $r$ the 
orbital separation between its constituents), as by the virial theorem 
$v^2\sim G_N M/r$.

The approach to solving for the dynamics of the two body problem adopted here
relies on the non-relativistic formulation of GR originally proposed in
\cite{Goldberger:2004jt}, see also \cite{Goldberger:2007hy} for a review, which
sets the problem in an Effective Field Theory (EFT) framework.

The use of field theory tools like Feynman diagrams in GR is not a novelty,
see e.g. \cite{Bertotti:1960xx,HariDass:1980tq,Damour:1995kt} for pioneer
work in this direction. With respect to these early works, the EFT approach
has the merit of recognizing scale separation as an organizational 
principle for systematic computation at the Lagrangian level.

Indeed the two body problem exhibits a clear separation of scales: the size of 
the  compact objects $r_s$, like
black holes and/or neutron stars, the orbital separation $r$ and the 
gravitational wave-length $\lambda$. Using again the virial theorem the 
hierarchy $r_s < r \sim r_s/v^2 < \lambda \sim r/v$ can be established.
The EFT approach allows to use the scale separation of the
physical problem to arrange a transparent and systematic power counting in the
expansion parameter, with physics at different scales related by
\emph{renormalization group flow}.
It has many common features with ordinary
quantum field theory (Feynman diagrams, divergence regularizations, logarithmic
running of physical observables) as both the \emph{classical} effective field 
theory described here and \emph{quantum} field theory share properties 
belonging to any \emph{field theory}.

The interest in the analytical description of gravitationally bound binary
systems has been revived in recent times by the activity of the above mentioned
GW observatories whose ouput is particularly sensitive to 
the GW phase. It is very important to have an accurate description of the 
waveform, whose shape depends on the source motion, for both maximizing the 
detection probability and for extracting the highest possible physical content
from candidate events.
Moreover data analysis techniques involve the generation of several tens of  
thousands to millions waveforms, thus requiring their analytical knowledge in 
order to have quick and efficient data analysis pipelines.

In particular the dynamical quantities allowing to determine physical
observables like the phase of the GW signal, are the energy of the bound orbit
and the emitted flux of GWs. Since signals falling in the detector band
sensitivity are in the very last stage of the coalescence, the binary system
orbits are expected to have circularized by then 
\cite{Peters:1963ux,MaggioreGW}, so the analytical quantities
to be computed are the energy of circular orbits and the 
gravitational flux as a function of the relative velocity of the binary 
constituents.

Moreover recent progress have made available numerical-relativity waveforms
emitted in the last $O(10)$ orbits of a binary system (including merger and 
ring-down) \cite{nrar}. In order to construct complete \emph{hybrid} waveforms 
encompassing all the stages of a coalescence from inspiral to merger and ring-down,
the highest possible accuracy on the analytical inspiral phase is necessary to
reduce the length of the numerically evolved part of the waveform, which is
in general very time consuming \cite{Ohme:2011zm}.

It is then expected that GW observations and numerical modeling will bring new 
inputs from both the phenomenological and the theoretical numerical side to the 
two body problem in General Relativity, whereas the effective field theory
approach described here is giving new momentum to the analytical studies on the 
theoretical side.

\section{General Theory}
\label{se:general}

The effective field theory approach to the GR two-body problem is analog to
other effective field theory approaches adopted to study specific systems in
particle physics, like the heavy quark field theory
\cite{Georgi:1990um,Isgur:1991wq}.
We want to study the dynamics of a pair of heavy and compact objects
(black holes/neutron stars) interacting through the exchange of gravitational
degrees of freedom and emitting GWs. 

The effective Lagrangian $\az_{ext}$
of \emph{any} extended object of size $r_{source}$ interacting
with a gravitational field with characteristic length-scale variation 
$L\gg r_{source}$, can be parametrized in terms of its mass $m$, spin tensor
$S_{ab}$ and higher order multipoles \cite{Goldberger:2005cd}
\footnote{We adopt the $(-,+,+,+)$ signature, $\tau$ is the 
proper time running along the source world-line, $u^\mu\equiv dx^\mu/d\tau$
is the 4-velocity of the center of mass.}
\be
\label{eq:ext}
\az_{ext}\supset
\int {\rm d}\tau\pa{-m-\frac 12 S_{ab}\omega_\mu^{ab}u^\mu\ +c_Q I_{ij} E^{ij}+c_J
J_{ij}B^{ij}+c_OI_{ijk}\dpa_{i}E_{jk}+\ldots}\,,\nonumber\\
\ee
where $\omega^{ab}_\mu$ is the spin connection coupling to the total 
angular momentum, while the electric (magnetic) tensor $E_{ij}$ ($B_{ij}$) is
defined by
\be
\ba{rcl}
\ds E_{ij}&=&\ds C_{\mu i\nu j}u^\mu u^\nu\,,\\
\ds B_{ij}&=&\ds \epsilon_{i\mu\nu\rho}\,u^\rho
C^{\mu\nu}_{\ \ j\sigma}\,u^\sigma\,,
\ea
\ee
decomposing the Weyl tensor $C_{\mu\alpha\nu\beta}$ analogously to the
electric and magnetic decomposition of the standard electromagnetic tensor 
$F_{\mu\nu}$.
This amounts to decompose the source motion in terms of the
world-line of its center of mass and moments describing its internal
dynamics. The $I_{ij}$, $I_{ijk}$, $J_{ij}$ tensors are the 
lowest order in an infinite series of source moments, the $2^{\rm n-th}$ electric 
(magnetic) moment in the above action scale at leading order as 
$mr_{source}^n$ ($mvr_{source}^n$), and they couple to the Taylor
expanded $E_{ij}$ ($B_{ij}$) which scales as $L^{-(1+n)}$, showing that the above 
multipole expansion is an expansion in terms of $r_{source}/L$.

Note that the multipoles, beside being intrinsic, can also be induced by the 
tidal gravitational field or by the intrinsic angular momentum (spin) of the
source. 
For quadrupole moments, the intrisic case will be explicitly dealt with in 
subsec.~\ref{sse:conspin}, whereas the tidal induced quadrupole moments 
$I_{ij},J_{ij}|_{tidal}\propto E_{ij},B_{ij}$ give rise to the following terms in the
effective action
\be
\label{eq:Spp}
\az_{tidal}=\int {\rm d}\tau \paq{c_EE_{ij}E^{ij}+c_B B_{ij}B^{ij}}\,.
\ee
This is also in full analogy with electromagnetism, where
for instance particles with no permanent electric dipole experience a quadratic
coupling to an external electric field.
Eq.~(\ref{eq:Spp}) can be used to describe a single, spin-less 
compact object in the field of its binary system companion. Considering that 
the Riemann tensor generated at a distance $r$ by a source of mass $m$ goes as
$m/r^3$, the finite size 
effect given by the $E_{ij}E^{ij}$ term goes as $c_E m^2/r^6$. For
dimensional reasons $c_E\sim G_N r_{source}^5$
\cite{Goldberger:2004jt}, thus showing that the finite size effects of a 
spherical symmetric body in the binary potential are $O(Gm/r)^5$ times the 
Newtonian potential, a well known result which goes under the name of
\emph{effacement principle} \cite{effacement} (the coefficient $c_E$ actually 
vanishes for black holes in $3+1$ dimensions \cite{Damour:1983cE,Kol:2011vg}).

One may consider the inclusion in $S_{ext}$ of monopole terms linear in
curvature invariants like $c_R\int d\tau R$ and 
$c_V\int R_{\mu\nu}\dot x^\mu\dot x^{\nu}$. 
However these terms can be safely omitted as they vanish by the Einstein 
equations outside the source generating them\footnote{Such terms
give $\delta$-like, unobservable contributions to the classical potential.
Equivalently, it can be shown that the field redefinition 
$g_{\mu\nu}\to g_{\mu\nu}+\delta g_{\mu\nu}$ with
\bdm
\delta g_{\mu\nu}=\int {\rm d}\tau\frac{\delta(x^\alpha-x^\alpha(\tau))}{\sqrt{-g}}
\paq{\pa{-c_R+\frac{c_V}2}g_{\mu\nu}-c_V u_\nu u_\nu}
\edm
can be used to set to zero the above terms linear in the curvature, see
\cite{Goldberger:2007hy} for details.}. As linear terms in the Ricci tensor and
Ricci scalar cannot appear, the terms involving the least number of derivatives
are the ones written above in eq.~(\ref{eq:ext}), in terms of the (traceless 
part of the) Riemann tensor.

\begin{figure}
\begin{minipage}[htb]{.35\linewidth}
\begin{center}
\includegraphics[width=\linewidth]{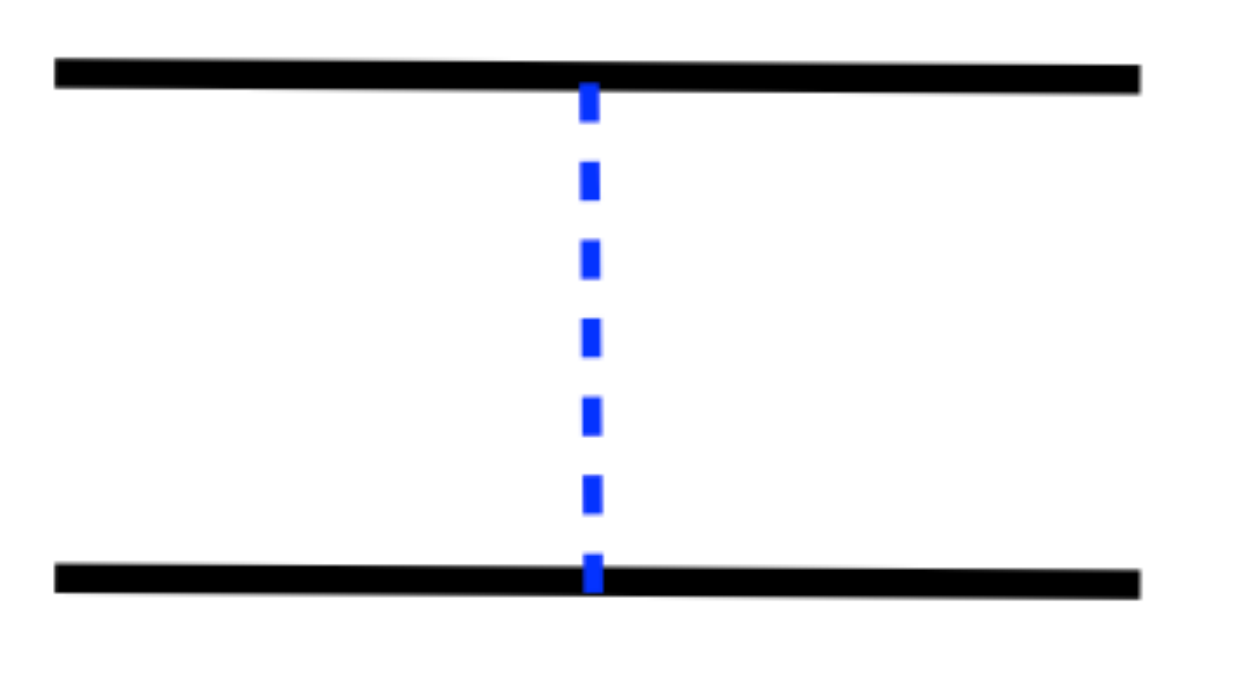}
\end{center}
\caption{Feynman graph accounting for the Newtonian potential.}
\label{fi:NewtPot}
\end{minipage}
\begin{minipage}{.63\linewidth}
\begin{center}
\includegraphics[width=.49\linewidth]{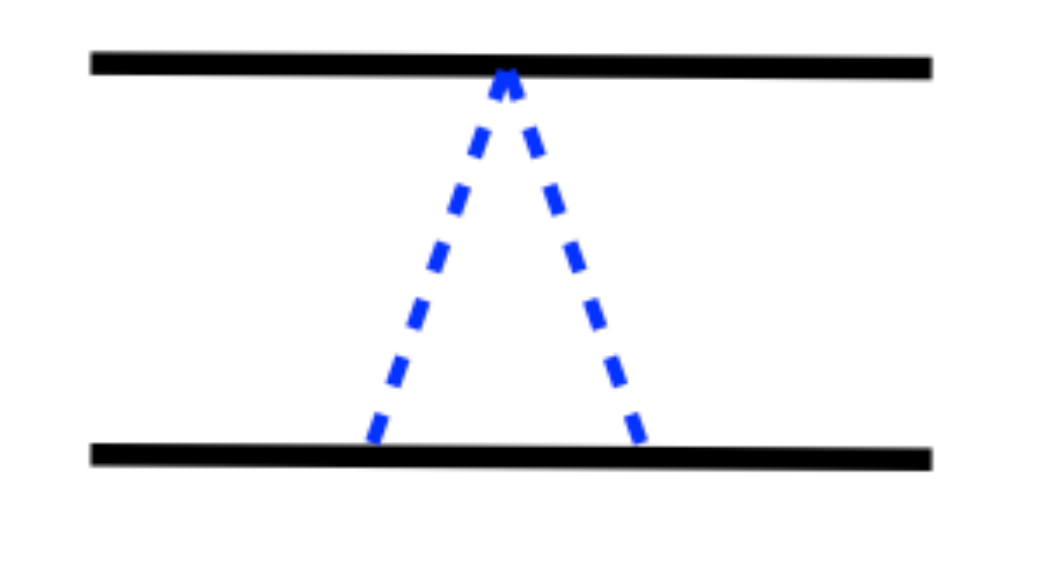}
\includegraphics[width=.46\linewidth]{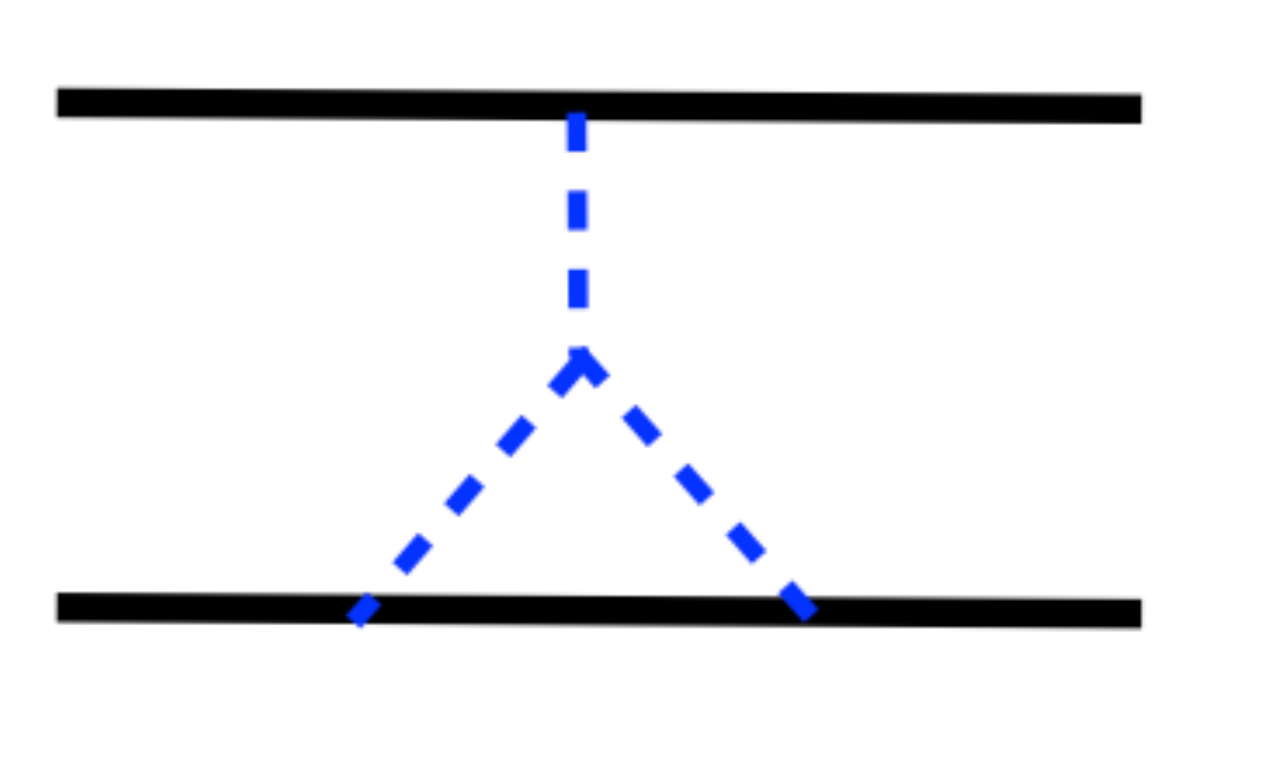}
\end{center}
\caption{Corrections to the Newtonian potential due to gravity non-linearities.
The left (right) diagram starts contributing at the 1(2)PN order.}
\label{fi:3phi}
\end{minipage}
\end{figure}

We will focus in the next section onto the derivation of the
effective potential of a binary system, obtaining the general relativistic
realization of the Newtonian potential.
This is obtained by \emph{integrating out} the degrees of freedom
that mediate the gravitational attraction to obtain a Fokker-type action 
$\az_{eff}$ describing the instantaneous interaction between objects 
parametrized by world-lines $x_{A,B}$. Formally this is achieved by
computing the Feynman path integral
\be
\label{eq:Feyn}
e^{i\az_{eff}}=\int \mathcal{D}h_{\mu\nu}\, e^{i\paq{S_{bulk}(\eta_{\mu\nu}+h_{\mu\nu})+
S_{ext}(x_a,\eta_{\mu\nu}+h_{\mu\nu})}}\,,
\ee
where $S_{bulk}$ involves only the gravitational degrees of freedom and is
given by the standard Einstein Hilbert action plus the harmonic gauge fixing 
term corresponding to the harmonic gauge used in \cite{lrr-2006-4}
\be
S_{bulk}=S_{EH}+S_{GF}\,,\quad S_{GF}\equiv-\Lambda^2\int {\rm d}t\,{\rm d}^dx\sqrt{-g}
\Gamma^\mu \Gamma_\mu\,,\quad\Gamma^\mu\equiv\Gamma^\mu_{\alpha\beta}g^{\alpha\beta}\,,
\ee
with $\Lambda=(32\pi G_N)^{-1/2}$ in $d=3$.

Because of the non-linearities of the Einstein Hilbert action and of the
gravity-matter coupling,
the functional integral in eq.~(\ref{eq:Feyn}) cannot be performed exactly,
but only perturbatively. 
For instance the leading perturbative order is represented 
by the diagram in fig.~\ref{fi:NewtPot}: it accounts for the potential
generated by the exchange of a gravitational degree of freedom.
Performing the above functional integral is equivalent to solving the non-linear
equations iteratively around the linear solution at the level of the action: 
the perturbative solution
can then be organized in Feynman diagrams as it is customary done in quantum 
field theory, however we stress once again that no quantum effects will be 
considered here.
 
In order to show in practice how the iterative solutions can be used to
efficiently generate the dynamics of the problem, we find convenient to 
decompose the metric in the form
\be
\label{met_nr}
g_{\mu\nu}=e^{2\phi/\Lambda}\pa{
\ba{cc}
-1 & A_j/\Lambda \\
A_i/\Lambda &\quad e^{- c_d\phi/\Lambda}\gamma_{ij}-
A_iA_j/\Lambda^2\\
\ea
}\,,
\ee
with $\gamma_{ij}=\delta_{ij}+\sigma_{ij}/\Lambda$,
$c_d=2\frac{(d-1)}{(d-2)}$, according
to the metric ansatz proposed in \cite{Kol:2007bc,Kol:2007rx} and 
reminiscent of the one first used in \cite{Blanchet:1989ki}.
On the previous ansatz $\az_{bulk}$ reduces to
\be
\az_{bulk}\simeq\int {\rm d}t\, {\rm d}^dx\pa{-c_d(\dpa\phi)^2+\ldots}\,,
\ee
where only the kinetic term of the gravitational field $\phi$ has been 
explicitly written.
The $\phi$ coupling to the source, which is implicit
from eq.~(\ref{eq:ext}), is $m/\Lambda\int d\tau\phi\,(1+{\cal O}(v^2))$ and 
neglecting all interaction terms of the field $\phi$, the Gaussian integration 
over $\phi$ in eq.~(\ref{eq:Feyn}) can be done exactly and leads to
\be
\label{eq:effG}
\!\!\!\!\!\!\!\!\!\!\!\!\!\!\!\!
S_{eff}=-m_1\int {\rm d}\tau_1 -m_2\int {\rm d}\tau_2+i\frac{m_1m_2}{2c_d\Lambda^2}
\int {\rm d}\tau_1{\rm d}\tau_2\,G(\tau_1-\tau_2,x_1(\tau_1)-x_2(\tau_2))\,,
\ee
where $G(t,x)$ is the Feynman Green function
\be
G(t,x)=-i\int \frac{{\rm d}k_0}{2\pi}\int_{\bf k}e^{-ik_0t+i{\bf k}\cdot x}
\frac 1{{\bf k}^2-k_0^2-i \epsilon}\,,
\ee
with $\int_{\bf k}\equiv\int \frac{{\rm d}^dk}{(2\pi)^d}$. It is now crucial to
take the non-relativistic limit 
in order to work at a given order in $v$. This is achieved by observing that 
the wave-number $k^\mu\equiv(k^0,{\bf k})$ of the gravitational modes 
mediating this interaction have $(k^0\sim v/r,k \sim 1/r)$, so in order to
have manifest power counting it is necessary to Taylor expand the propagator
\be
\label{eq:propT}
\ba{rcl}
\ds G(t,x)&\simeq &\ds -i\int \frac{{\rm d}k_0}{2\pi}\int_{\bf k}
e^{-ik_0t+i{\bf k}\cdot x} \frac 1{{\bf k}^2}\pa{1+\frac{k_0^2}{{\bf k}^2}+\ldots}\\
&=&\ds-i\int_{\bf k}
\delta(t_0)e^{i{\bf k}\cdot x}\frac 1{{\bf k}^2}\pa{1-\frac{\dpa_t^2}{{\bf k}^2}+\ldots}\,,
\ea
\ee
where in the last passage the integral over $k_0$ has been performed explicitly
after trading the $k_0$ factors for time derivative operators.
Note that we did not write any $i \epsilon$ term in eq.~(\ref{eq:propT}) as, in the
non-relativistic kinematical region we are interested in here, where the gravitational
mode cannot be on-shell, the pole prescription is inessential.
The individual particles can also exchange \emph{radiative} gravitons (with
$k_0\simeq k\sim v/r$), but such processes give sub-leading contributions to the
effective potential in the PN expansion, and they will be dealt with in 
subsec.~\ref{sse:radreac}. In other words we are not integrating out the entire
gravity field, but the specific off-shell modes in the kinematic region 
$k_0\ll k$.

\begin{figure}
  \begin{minipage}[htb]{.33\linewidth}
    \begin{center}
      \includegraphics[width=\linewidth]{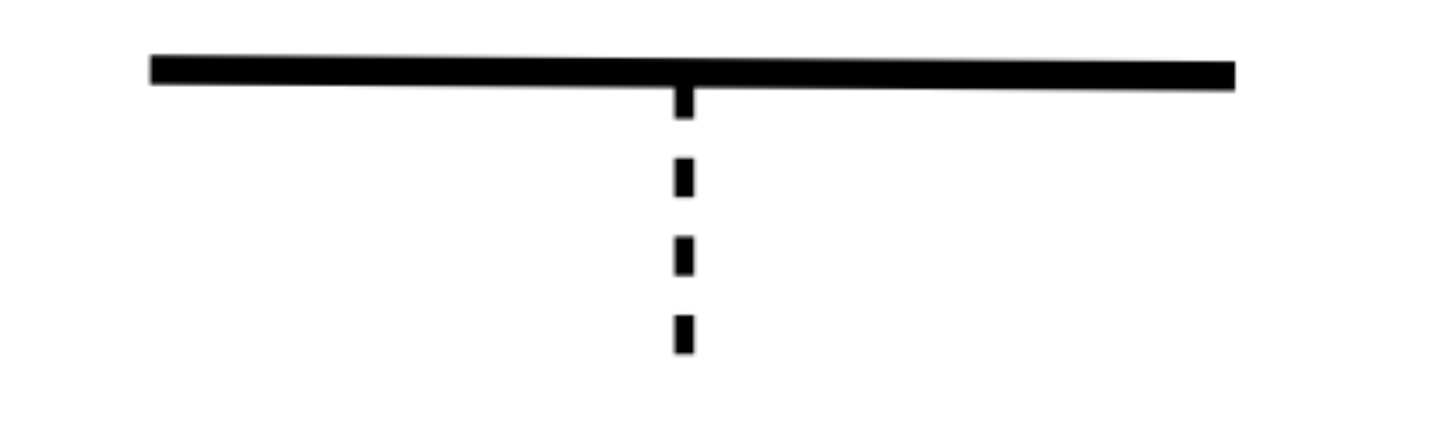}
      \caption{Vertex scaling:\\$\ds\frac m\Lambda dt d^dk\sim \frac m\Lambda
        \frac {r^{1-d}}v$}
      \label{fi:rul1}
    \end{center}
  \end{minipage}
  \begin{minipage}{.33\linewidth}
    \begin{center}
      \includegraphics[width=\linewidth]{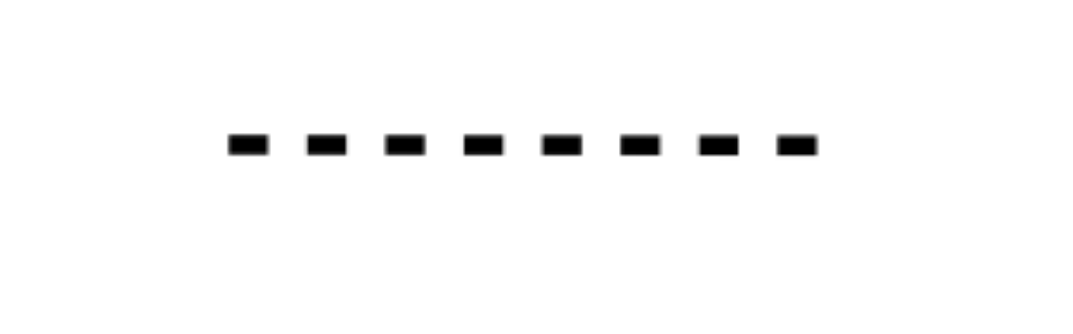}
      \caption{A Green function is represented by a propagator, with scaling:\\
        $\delta(t)\delta^d(k)/k^2\sim vr^{1+d}$}
      \label{fi:rul2}
    \end{center}
  \end{minipage}
  \begin{minipage}{.33\linewidth}
    \begin{center}
      \includegraphics[width=\linewidth]{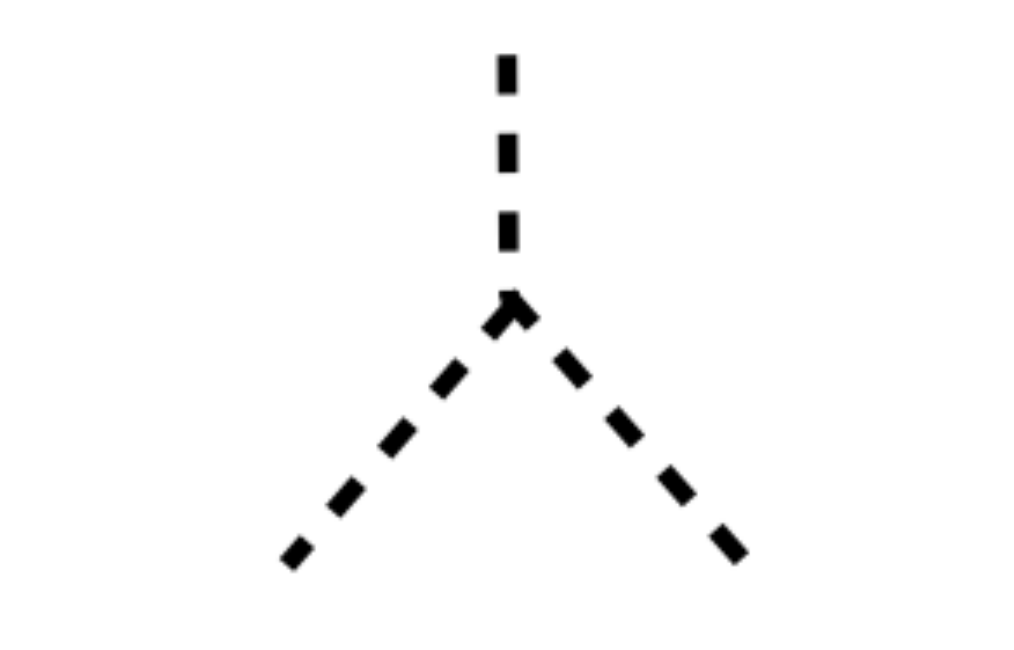}
      \caption{Triple internal vertex scaling:\\
        $\ds\frac{(k^2,kk_0,k_0^2)}\Lambda\delta^{d}(k)dt(d^dk)^3\sim
        \frac{(1,v,v^2)}{r^{1+2d}v\Lambda}$}
        \label{fi:rul3}
    \end{center}
  \end{minipage}
\end{figure}

We are aiming at computing
an effective action giving the correct action-at-distance, with retardation
effects taken into account by the Taylor expansion in eq.~(\ref{eq:propT}).
After substituting in the effective action (\ref{eq:effG}) the explicit form
of the source-gravity coupling one obtains
\be
S_{eff}\supset m_Am_B\int {\rm d}t\pa{1+{\cal O}(v_1^2)}\pa{1+{\cal O}(v_2^2)}
\int_{\bf k}\frac{e^{i {\bf k} \cdot r}}{{\bf k}^2}\pa{1-\frac{\dpa_t^2}{{\bf k}^2}+
\ldots}
\,.
\ee
Once the time derivatives act on the exponential they will introduce velocity
dependent terms in the effective action, so in order to have a consistent
calculation at any given order in $v^2$, we
have to remember the virial theorem, which makes diagrams of the type in 
fig.~\ref{fi:3phi} also potentially of order $v^2$ with respect to the leading
one in fig.~\ref{fi:NewtPot}.
The power counting of diagrams can be made systematic by the rules given in
fig.~\ref{fi:rul1},\ref{fi:rul2},\ref{fi:rul3}, which can be generalized to 
higher order interaction vertices.

Intermediate massive object lines, (like the ones in fig.~\ref{fi:3phi})
have no propagator associated, as they represent a static source (or sink) of 
gravitational modes. At the graviton-massive object vertex momentum is 
\emph{not} conserved, as the graviton momentum is ultra-soft compared to the
massive source.
E.g. in the diagram of fig.~\ref{fi:NewtPot}, where the massive object emits a 
single gravitational mode, it recoils by a fractional amount $dp/p$ roughly 
given by $dp/p=\hbar k/(mv)\sim \hbar/L$, being $L=mrv$ the macroscopic angular
momentum of the binary system. For the phenomenological application we are 
aiming at, $\hbar/L\sim 10^{-77}(M/M_\odot)^{-2}(v/0.1)$ is ridiculously small and
completely negligible.

\begin{figure}
  \begin{minipage}[htb]{.49\linewidth}
    \begin{center}
      \includegraphics[width=\linewidth]{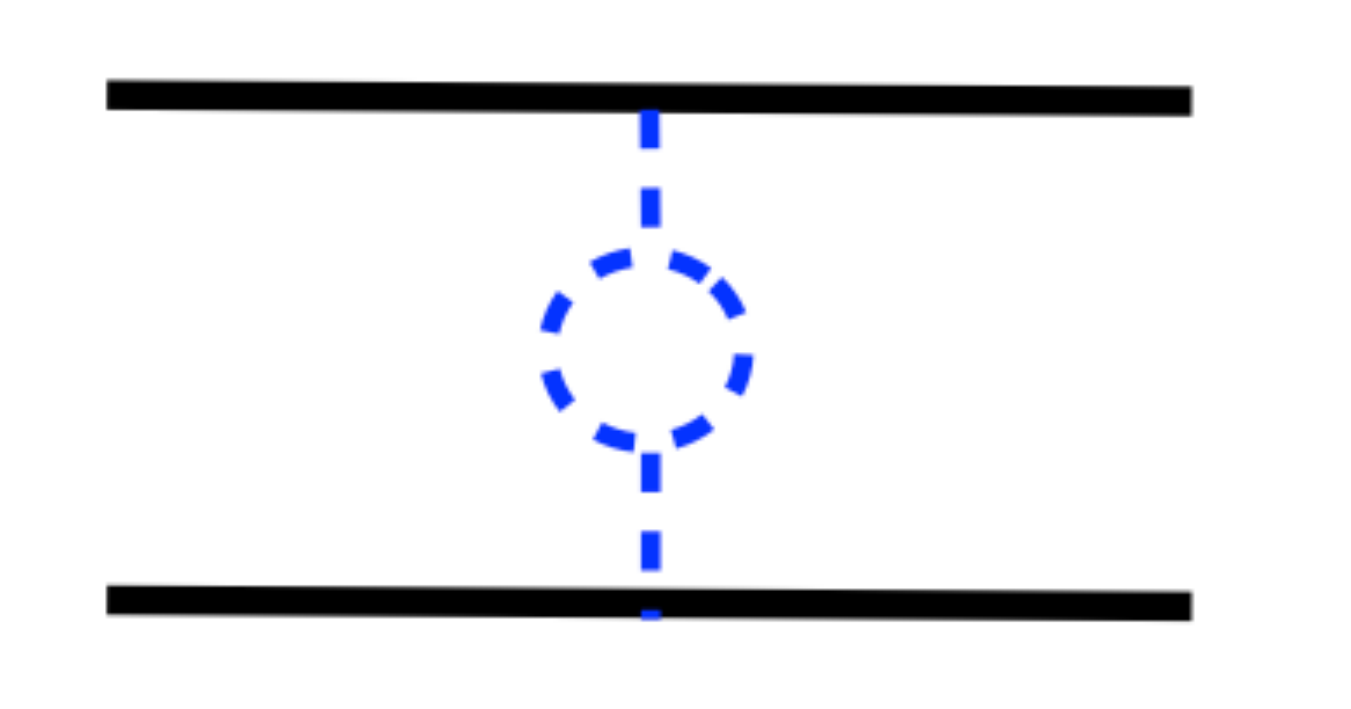}
      \caption{Quantum contribution to the 2-body potential.}
      \label{fi:loop}
    \end{center}
  \end{minipage}
  \begin{minipage}{.49\linewidth}
    \begin{center}
    \includegraphics[width=.9\linewidth,angle=180]{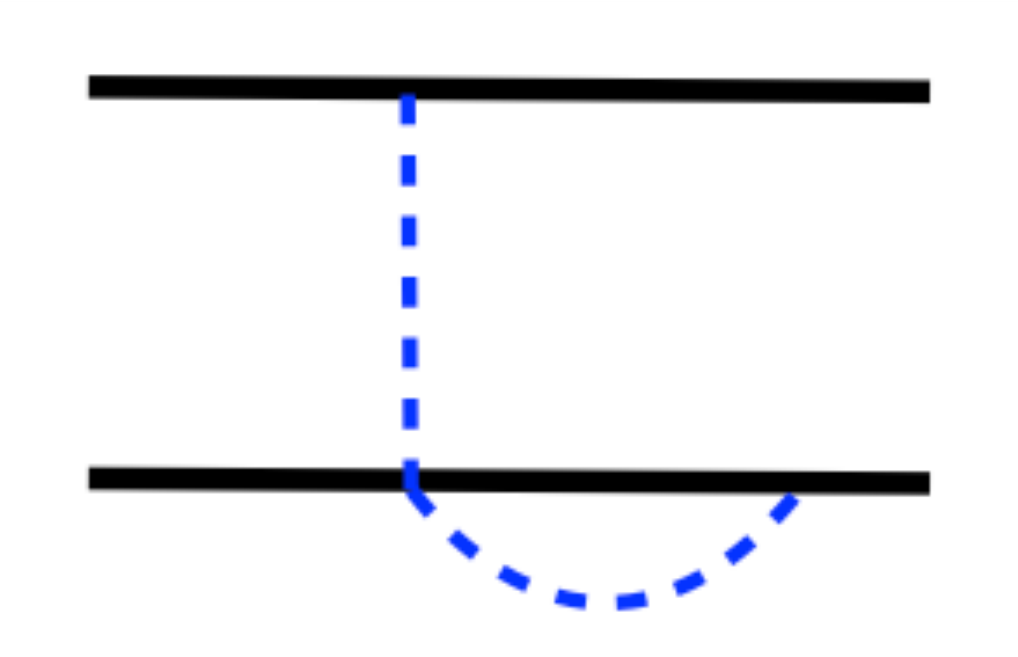}
    \caption{Diagram giving a power-law divergent contribution to the mass.}
    \label{fi:corr0}
    \end{center}
  \end{minipage}
\end{figure}

Consistently with neglecting any quantum effect, diagrams like the one
in fig.~\ref{fi:loop} will not be considered. Even explicitly restoring $\hbar$ 
into the 
definition of the path integral in eq.~(\ref{eq:Feyn}) to establish the correct 
dimensions of the exponential term, after evaluating only diagrams at tree 
level (in the quantum language) all physical result will be $\hbar$-independent.
According to the standard rules for taking into account powers of $\hbar$
involved in Feynman diagrams, each vertex brings in an inverse power of $\hbar$
and each internal line a power of $\hbar$, making the quantum scaling of
diagram accounted $\hbar^{I-V}=\hbar^{L-1}$, using the standard relationship
$L=I-V+1$ among number of loops $L$, vertices $V$ and propagators $I$. Applying
this power counting rule to the graph in fig.~\ref{fi:loop}, say, shows that it
scales
as $\hbar/L$ with respect to the Newtonian potential, so it is completely 
negligible.\footnote{Note that adopting a quantum field theory description of a 
second quantized massive particle $\psi$ coupled to gravity would lead to the
same result as here, once the non-relativistic limit is taken, see
\cite{Donoghue:1994dn}.}

After integrating out the potential graviton we will be left with an effective
action where some of the original operators will be renormalized and new, 
local ones will be generated in infinite numbers (but finite at each PN order),
with the coefficients of the generated operators being the 
\emph{Wilson coefficients}.
Note that some graphs will be actually divergent
like the one in fig.~\ref{fi:corr0}, which gives a divergent contribution to
the effective potential
\be
\label{eq:corr0}
fig.~\ref{fi:corr0}\simeq \frac{G_N^2m_1^3m_2}r\int_{\bf k}
\frac 1{{\bf k}^2}\,.
\ee
Actually graphs like this can be consistently discarded,
and indeed vanish in dimensional regularization, as an effective theory
is not supposed to correctly portrait the full theory at arbitrary high
energy scales. Divergences like the one of eq.~(\ref{eq:corr0}) can be
accounted for by shifting the input parameters in the starting Lagrangian
(like the mass of the binary constituents), as we are not aiming at 
\emph{predicting} those parameters, but just take them as inputs (see 
\cite{Rothstein:2003mp} for a thorough discussion along this line).
We shall discuss in the following sections three other kinds of divergence, 
associated to $O(3{\rm PN})$ gauge artifacts, to long-distance effects and to
short-distance (or ultra-violet) incompleteness of the effective theory.

As it is standard in perturbative field theory calculations,
diagrams contributing to the effective action are only the \emph{connected}
ones, i.e. those in which
following Green function's lines all the vertices can be connected.

The effective theory at the orbital case, in the spin-less case, can treat
the binary constituents as point-like until 5PN order, as this is the order
at which finite size effects
come into play, so the theory can be consider ultra-violet (UV) complete up to 
that order (the finite size spinning effects will be discussed in 
subsec.~\ref{sse:conspin}).

In sec.~\ref{se:radiation} we shall consider the effective action of eq.~(\ref{eq:ext})
to describe the binary system as a single extended object coupled to 
gravity in order to compute observables related to the emission of GWs. 
The starting point will be the action in eq.~(\ref{eq:ext}), where
the first two terms will not be responsible for radiation, as at leading order
they couple the gravitational modes to the conserved mass monopole and to the
total angular momentum.

In order to have full predictive power, the effective theory in terms of the
multipole moments at the orbital scale
will have to be matched to the
theory at the orbital scale in order to express the binary multipoles in terms
of individual constituent parameter.
It will turn out that in computing the radiation back-reaction on the source
at the scale $\lambda\sim r/v$, a logarithmic divergence will appear,
showing the UV incompleteness already at
$v^3$, and requiring that the singularity be resolved by considering
the theory at the smaller, orbital scale (in the calculation of the
emitted flux the incompleteness will appear at $v^6$ order).

While it is possible to absorb power-divergences into bare parameters of the
original Lagrangian, 
as it is usual in field theory, logarithmic divergences will introduce a 
spurious 
dependence on an arbitrary scale $\mu$: in order to cancel the $\mu$ dependence
from physical observables, a compensating dependence of the input parameters has
to be imposed, leading to a fully classical implementation of the 
\emph{renormalization group} equation, implying that physical parameters running
with $\mu$ will take different values when probed at different length scale, as
it will be explicitly shown in subsec.~\ref{sse:radreac}.

\section{Conservative}
\label{se:cons}
The conservative dynamics of a binary system involves processes characterized by no incoming nor outgoing radiation:
in diagrammatic terms, this means absence of external radiative graviton lines.
{\em Internal} radiative propagators (meaning that the radiation graviton is emitted and then reabsorbed by the system)
can in principle be present and indeed appear at 4PN, giving rise to the so called {\em tail terms} studied in \cite{Blanchet:1987wq,Blanchet:1993ng,Blanchet:1993ec};
we will deal with this peculiar effect in subsec.~\ref{sse:radreac}, while restricting the discussion of this section to diagrams involving potential gravitons only.

The GW length $\lambda$ thus being irrelevant at this stage, the only scales of the problem are the size of the stars/black holes $r_s$ and the orbital radius 
$r$.
The main goal here is to determine the dynamics of the system as a function of the orbital parameters and of the internal features of the stars, such as mass and spin, and other (like $c_{E,B}$) which appear as Wilson coefficients to be 
fixed by a matching procedure at the scale $r_s$.

The general strategy consists in
\begin{enumerate} 
\item writing down all the relevant vertices of the effective theory and 
determine their $v^2$ and $G_N$ scaling 
\item building all the Feynman diagrams which are
relevant to the desired PN order
\item computing the Feynman integrals by Taylor expanding potential graviton
propagators around $k_0=0$\,.
\end{enumerate}
Power-law divergences arising at this point are automatically reabsorbed by 
dimensional regularization, while the logarithmic divergences appearing for 
the first time at 3PN can be eliminated by means of a world-line 
re-parametrization.

The pure gravity sector of the theory
can be expanded up to the desired order in terms of the Kaluza-Klein variables
introduced in eq.~(\ref{met_nr}).
We report here the expansion up to terms relevant at 4PN
\footnote{$\hat{\Gamma}^i_{jk}$ is the connection of the purely spatial metric $\gamma_{ij}$
, $F_{ij}\equiv A_{j,i}- A_{i,j}$ and indices must be raised and
contracted via the $d$-dimensional metric tensor $\gamma$; on the other hand
all the spatial derivatives are meant to be simple (not covariant) ones and,
when ambiguities might raise, gradients are always meant to act on
contravariant fields (so that, for instance,
$\vec{\nabla}\!\!\cdot\!\!\vec{A}\equiv\gamma^{ij}A_{i,j}$
and $F_{ij}^2\equiv\gamma^{ik}\gamma^{jl}F_{ij}F_{kl}$).} 
(see also \cite{Kol:2010si} for a derivation): 
\renewcommand{\arraystretch}{1.4}
\be
\label{bulk_action}
S^{4PN}_{bulk}&\simeq&\int {\rm d}t\, {\rm d}^dx\sqrt{-\gamma}
\left\{\frac{1}{4}\left[(\vec{\nabla}\sigma)^2-2(\vec{\nabla}\sigma_{ij})^2-\left(\dot{\sigma}^2-2(\dot{\sigma}_{ij})^2\right){\rm e}^{\frac{-c_d \phi}{\Lambda}}\right]\right.\nonumber\\
&&
- c_d \left[(\vec{\nabla}\phi)^2-\dot{\phi}^2 {\rm e}^{-\frac{c_d\phi}{\Lambda}}\right]
+\left[\frac{F_{ij}^2}{2}+\left(\vec{\nabla}\!\!\cdot\!\!\vec{A}\right)^2 -\dot{\vec{A}}^2 {\rm e}^{-\frac{c_d\phi}{\Lambda}} \right]
{\rm e}^{\frac{c_d \phi}{\Lambda}}\nonumber\\
&&
+2\frac{\left[F_{ij}A^i\dot{A^j}+\vec{A}\!\!\cdot\!\!\dot{\vec{A}}(\vec{\nabla}\!\!\cdot\!\!\vec{A})\right]
{\rm e}^{\frac{c_d \phi}{\Lambda}}-c_d\dot{\phi}\vec{A}\!\!\cdot\!\!\vec{\nabla}\phi}{\Lambda}-c_d\frac{\dot{\phi}^2\vec{A}^2}{\Lambda^2}\nonumber\\
&&
+2 c_d \left(\dot{\phi}\vec{\nabla}\!\!\cdot\!\!\vec{A}-\dot{\vec{A}}\!\!\cdot\!\!\vec{\nabla}\phi\right)
+\frac{\dot{\sigma}_{ij}}{\Lambda}\left(-\delta^{ij}A_l\hat{\Gamma}^l_{kk}+ 2A_k\hat{\Gamma}^k_{ij}-2A^i\hat{\Gamma}^j_{kk}\right)\nonumber\\
&&
-\left.\frac{1}{\Lambda}\left(\frac{\sigma}{2}\delta^{ij}-\sigma^{ij}\right)
\left({\sigma_{ik}}^{,l}{\sigma_{jl}}^{,k}-{\sigma_{ik}}^{,k}{\sigma_{jl}}^{,l}+\sigma_{,i}{\sigma_{jk}}^{,k}-\sigma_{ik,j}\sigma^{,k}
\right)\right\}\,.
\ee
\renewcommand{\arraystretch}{1.}

All the bulk vertices and propagators needed up to 4PN order can be derived from
eq.~(\ref{bulk_action}).
We write down explicitly the Green function expressions
in terms of the space Fourier-transformed variables
\be
\label{Fourk}
W^a_{\bf k}(t)  \equiv \ds\int {\rm d} t\, {\rm d}^dx\, W^a(t,x) e^{-i{\bf k}\cdot x}\,\quad
 {\rm with\ } W^a=\{\phi,A_i,\sigma_{ij}\}\,:
\ee
\be
\label{propagators}
\ds P[W^a_{\bf k}(t_a)W^b_{\bf k'}(t_b)]&=&\ds \frac{1}{2} P^{aa}\delta_{ab}
\ds (2\pi)^d\delta^{d}({\bf k}+{\bf k'}){\cal P}({\bf k}^2,t_a,t_b)\delta(t_a-t_b)\,,
\ee
with $P^{\phi\phi}=-\frac{1}{c_d}$, $P^{A_iA_j}=\delta_{ij}$,
$P^{\sigma_{ij}\sigma_{kl}}=-\left(\delta_{ik}\delta_{jl}+\delta_{il}\delta_{jk}+(2-c_d)\delta_{ij}\delta_{kl}\right)$ 
and
\be
\label{propexp}
{\cal P}({\bf k}^2,t_a,t_b)=\frac{i}{{\bf k}^2-\partial_{t_a}\partial_{t_b}}\simeq
\frac{i}{{\bf k}^2}\left(1+\frac{\partial_{t_a}\partial_{t_b}}{{\bf k}^2}+
\frac{\partial_{t_a}^2\partial_{t_b}^2}{{\bf k}^4}\dots\right)\,.
\ee
As desirable, the three polarization fields $\phi$, $A$, $\sigma$ do not mix at the quadratic level.

\begin{figure}
\includegraphics[width=.3\linewidth]{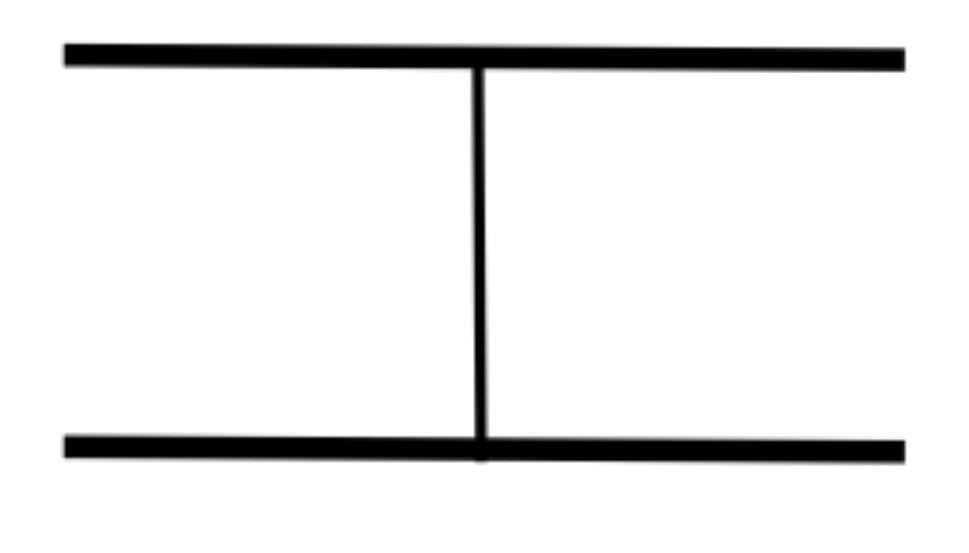}
\includegraphics[width=.7\linewidth]{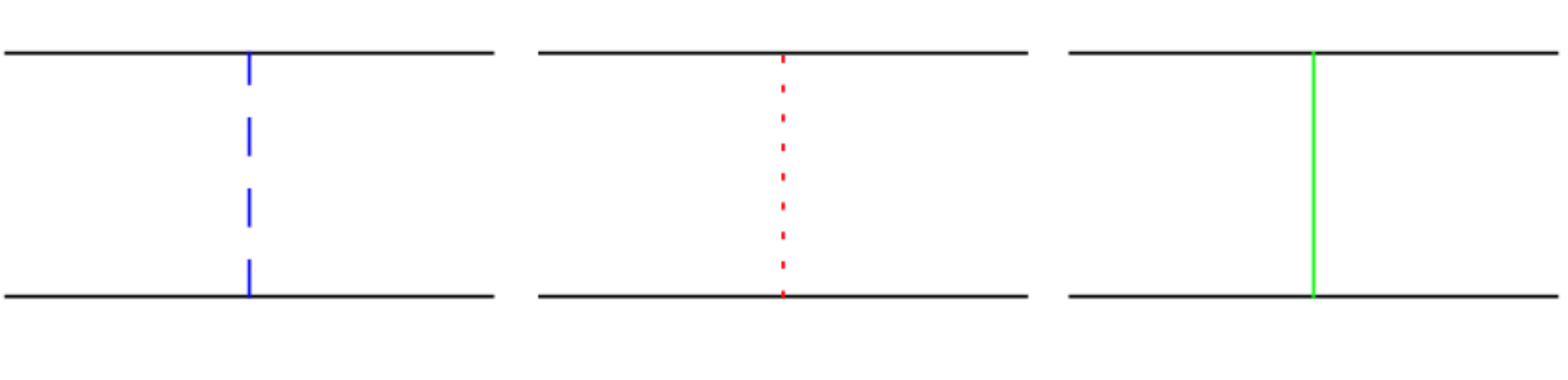}
\caption{The only topology contributing at order $G_N$, along with the three diagrams that are derived from it for spin-less objects.
The $\phi$, $A$ and $\sigma$ propagators are represented respectively by blue dashed, red dotted 
  and green solid lines.}
\label{diaG1}
\end{figure}

A convenient strategy for building all the Feynman diagrams has two steps (as first done in \cite{Gilmore:2008gq}).
At first one determines the shape of the diagram (henceforth called topology), which fixes the powers of $G_N$ through the following simple rules: 
a bulk $n$-vertex gives $G_N^{n/2-1}$, and a matter interaction vertex with $n$ graviton lines gives $G_N^{n/2}$, see e.g. figs.~\ref{fi:rul1},\ref{fi:rul3}.
Starting form the lowest order topology of fig.~\ref{diaG1},  all the higher order ones can be generated iteratively 
by adding a new propagator with one extremum on one of the two stars' word-lines, and the other to any other element
(a bulk vertex, a vertex located on the other star's word-line, or in the middle of an other propagator in order to create a new 3-vertex),
but not on the same word-line of the first extremum, since as pointed out in sec.~\ref{se:general},
topologies involving propagators
that start and end on the same star word-line as in fig.~\ref{fi:corr0} do not have to be considered.

Then, any given topology is ``filled" with the various $\phi,A,\sigma$ field 
propagators and with the different matter interaction vertices given by
eq.~\ref{eq:ext}: these elements determine the powers of $v$ characterizing the 
diagram, and thus its PN order.
The simplest example of this procedure is depicted in fig.~\ref{diaG1}.

The advantage of this procedure is that each topology is associated with a 
specific class of Feynman integrals,
so that all the diagrams belonging to the same topology can be computed using the same integration strategy.
Moreover, topologies which can be split into sub-topologies do not present any new difficulty from
the computational point of view because the corresponding amplitudes are given
by the product of the sub-topology ones which can thus be evaluated separately.

\subsection{The spin-less case}
Let us consider first the gravity-matter coupling for the non-spinning case and
postpone the more complicate spinning case to the next subsection.
In this case the finite size of the binary system components does not enter the 
dynamics until 5PN order (because of the effacement principle discussed in 
sec.~\ref{se:intro}), so the gravity-source
coupling reduces to the mass monopole term, which can be written as
\renewcommand{\arraystretch}{1.4}
\be
\label{matter_grav}
S_{pp}=-m\ds \int {\rm d}\tau = \ds-m\int {\rm d}t\ e^{\phi/\Lambda}
\sqrt{\pa{1-\frac{\vec{A}\!\cdot\!\vec{v}}{\Lambda}}^2
-e^{-c_d \phi/\Lambda}\pa{v^2+\frac{\sigma_{ij}}{\Lambda}v^iv^j}}\,,
\nonumber\\
\ee
\renewcommand{\arraystretch}{1.4}

We have now all the elements to complete step (ii), that is to determine all 
the relevant graphs at a given PN order.
The only diagram contributing at Newtonian level is the first one drawn next to the ${\cal O}(G_N)$ topology in fig.~\ref{diaG1},
because $\phi$ is the only polarization whose particle interaction vertex does not depend on $v$ at leading order.

\begin{figure}
\includegraphics[width=1\linewidth]{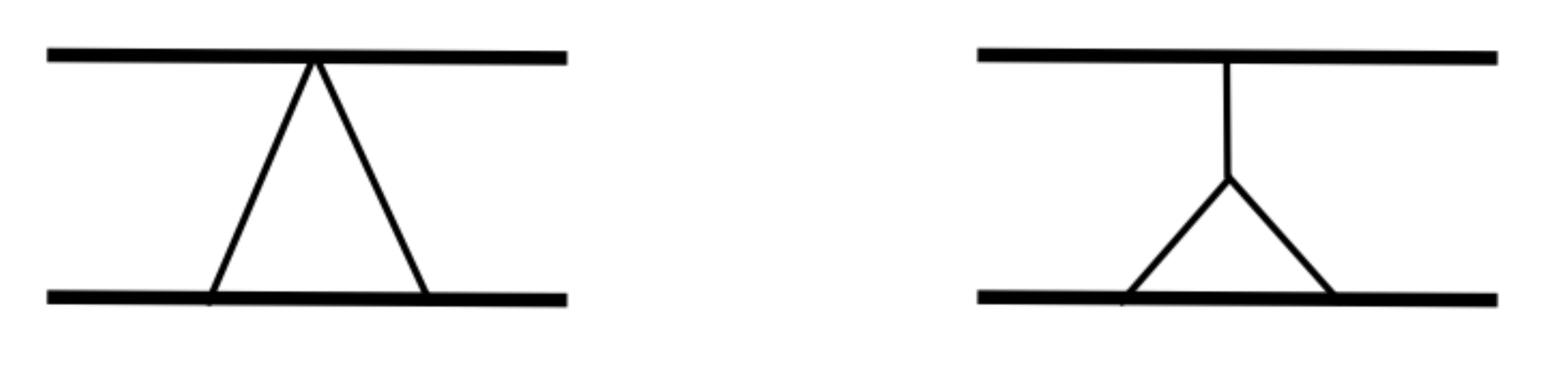}
\caption{The two $G_N^2$ topologies. The left one affects the dynamics with one diagram at 1PN, one more at 2PN, one more at 3PN and two more at 4PN; the one at the right with 7 diagrams at 2PN, 6 more at 3PN and 5 more at 4PN}
\label{topG2}
\end{figure}

The 1PN diagrams can scale ad $G_N v^2$ or $G_N^2$; in the first category fall the same diagram as before (which has to be
computed at ${\cal O}(v^2)$ by expanding the particle interaction vertices and the $\phi$ propagator according to 
eqs.~(\ref{matter_grav}) and (\ref{propexp}), respectively),
as well as the second diagram in fig.~\ref{diaG1}, which carries two powers of $v$ (one at each particle interaction vertex) at leading order.
As to the $G_N^2$ graphs, one has to consider the new topologies shown in 
fig.~\ref{topG2} and take the $v$-independent part of the Feynman diagrams.
Only the diagram with a $\phi^2$ source-gravity vertex contributes
at this PN order while the diagrams involving a triple bulk interaction vertex
can be discarded at this order because they carry at least two powers of $v$\footnote{The only diagram that does not pay any $v$ penalty factor at the 
particle-gravity interaction vertices is the one with three $\phi$'s, but the 
$\phi^3$ bulk interaction vertex, as it can be seen in eq.~(\ref{bulk_action}),
carries two time derivatives, giving two powers of $v$ in the final 
amplitude.}.
Note that the surviving $G_N^2$ diagram at 1PN order is clearly factorisable 
in terms of two ``Newtonian" topologies, so its calculation is 
straightforward.

\begin{figure}
\includegraphics[width=1\linewidth]{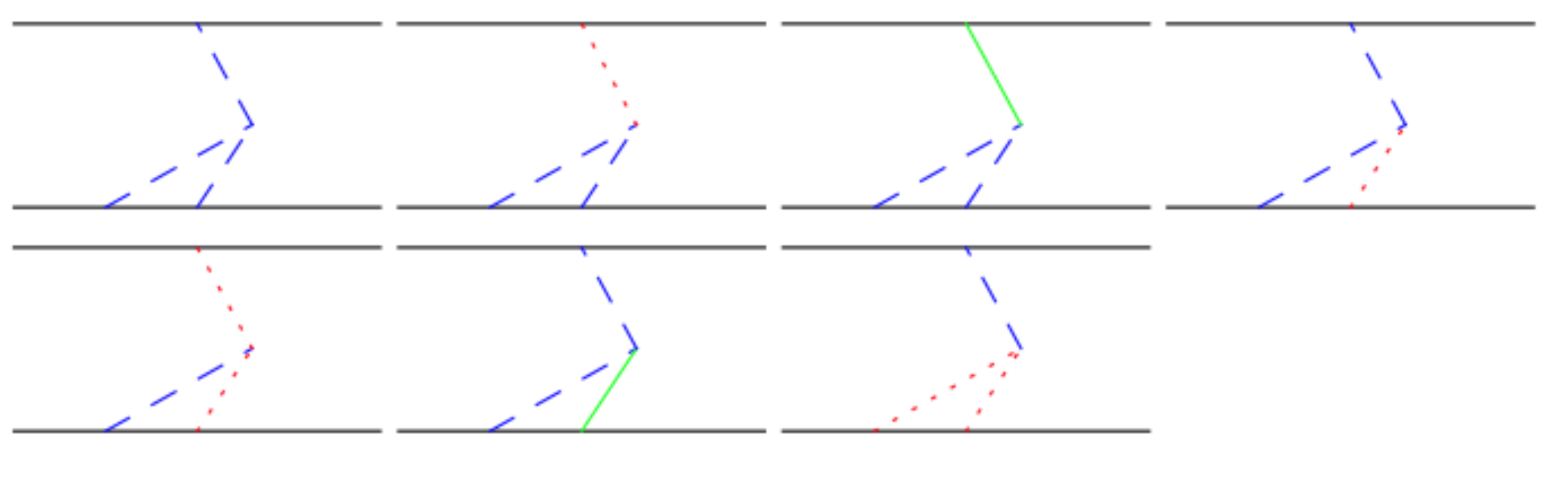}
\caption{The 2PN diagrams coming from the right $G_N^2$ topology of fig.~\ref{topG2}.}
\label{diaG2PN2}
\end{figure}

\begin{figure}
\includegraphics[width=1\linewidth]{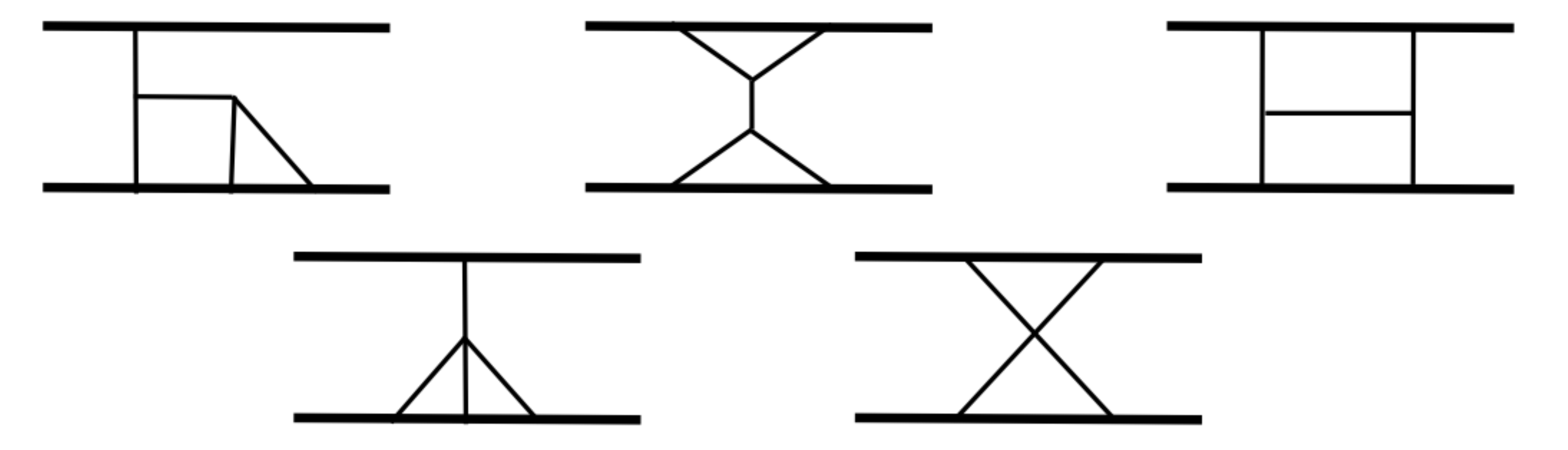}
\caption{The 5 irreducible $G_N^3$ topologies: the upper 3 are relevant already at 2PN, while the lower 2 at 3PN. There are other 4 $G_N^3$ topologies not shown here (2 of which are relevant at 2PN and 2 relevant at 3PN) as they are simple
products of lower order ones.}
\label{topG3}
\end{figure}

At 2PN we have to consider, as well as the previously analyzed diagrams with 
the appropriate factors of $v$ from the expansion of the propagators and vertices,
also several new diagrams generated by the 
topologies already considered (see fig.~\ref{diaG2PN2} for an example),
as well as the ones generated by brand new, $G_N^3$ topologies.
At 2PN, 5 of them are relevant (each one providing a single diagram), 2 of which
being merely trivial compositions of three ``Newtonian" topologies.
The 3 irreducible ones are shown in the upper part of fig.~\ref{topG3}.
The first computations of the effective 2PN Lagrangian within the EFT framework have been done in \cite{Gilmore:2008gq},
using the same Kaluza-Klein decomposition adopted here, and in \cite{Chu:2008xm}.

We conclude the topology and diagram classification before moving to amplitudes calculation.
At 3PN 63 new diagrams have to be considered, and 6 of them come from the two topologies shown in the lower part of fig.~\ref{topG3}, which are the only new irreducible topologies needed at this order:
in particular, all the 8 $G_N^4$ topologies needed at 3PN are factorisable in terms of simpler ones.
The 3PN calculation within the EFT framework has been performed in \cite{Foffa:2011ub} by means
of a semi-automated algorithm thus making the EFT technique match what was the state of the art at the time in this sector
of the theory.

\begin{figure}
\includegraphics[width=1\linewidth]{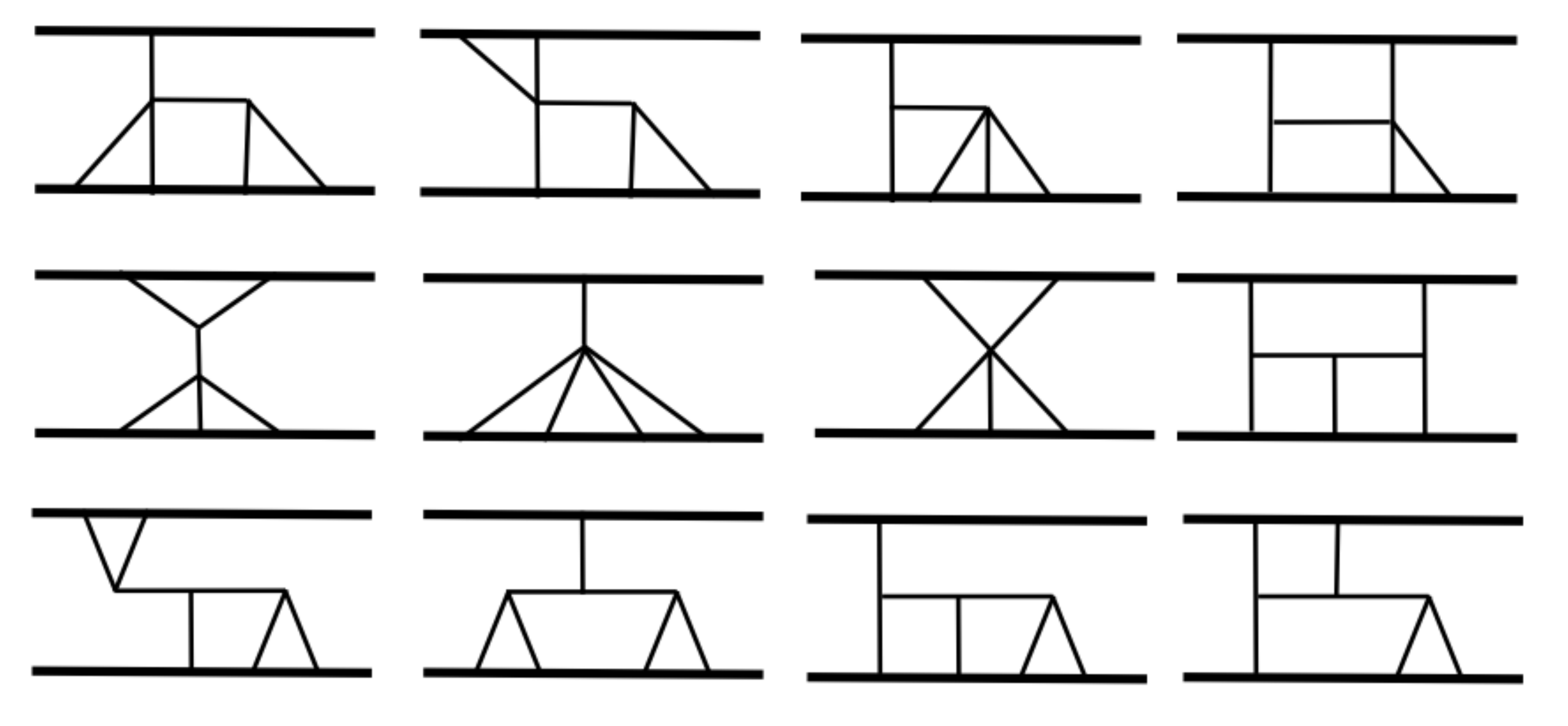}
\caption{The 12 irreducible $G_N^4$ topologies. They all give contribution to the 4PN dynamics.}
\label{topG4}
\end{figure}

\begin{figure}
\includegraphics[width=.61\linewidth]{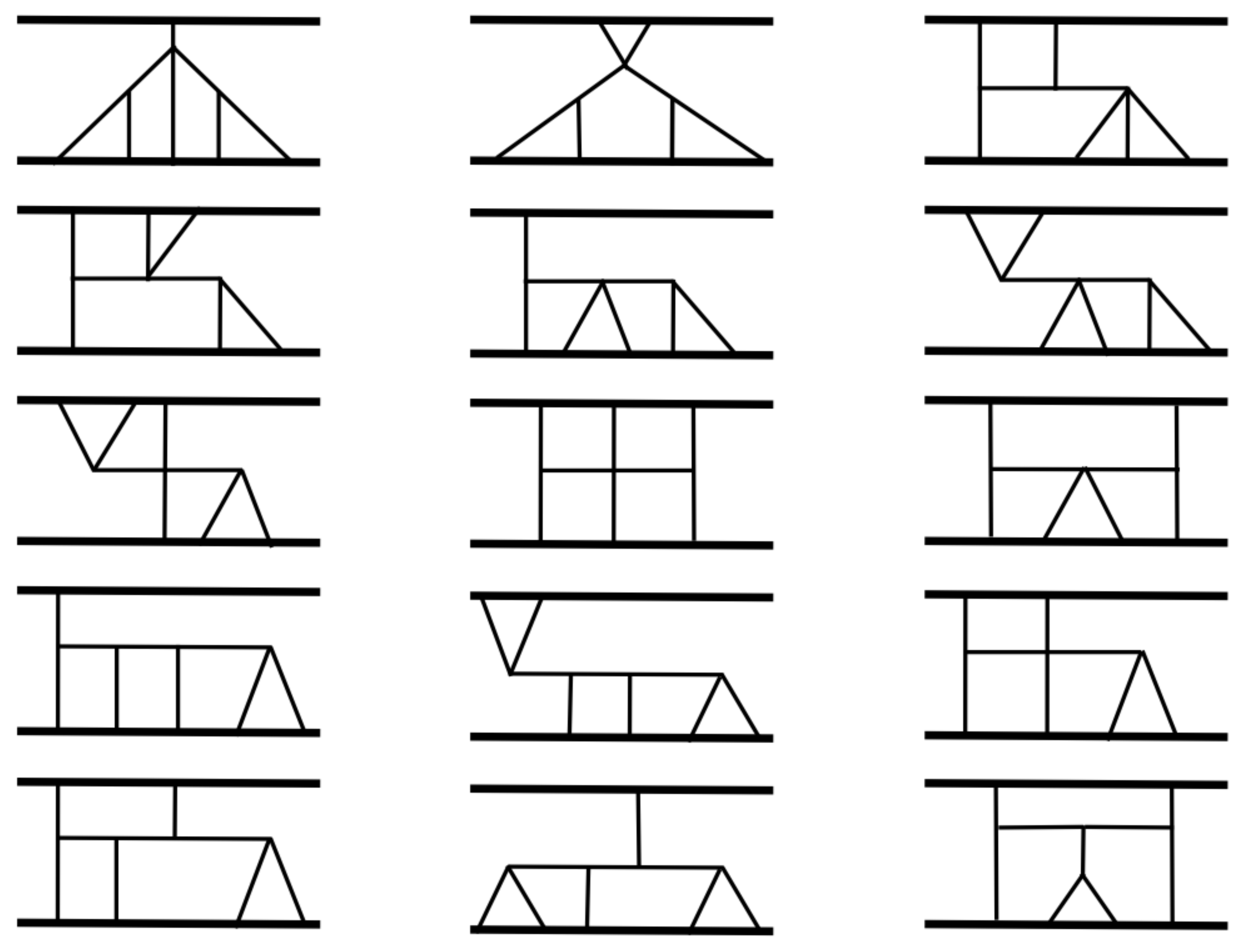}
\hspace{.3cm}
\includegraphics[width=.39\linewidth]{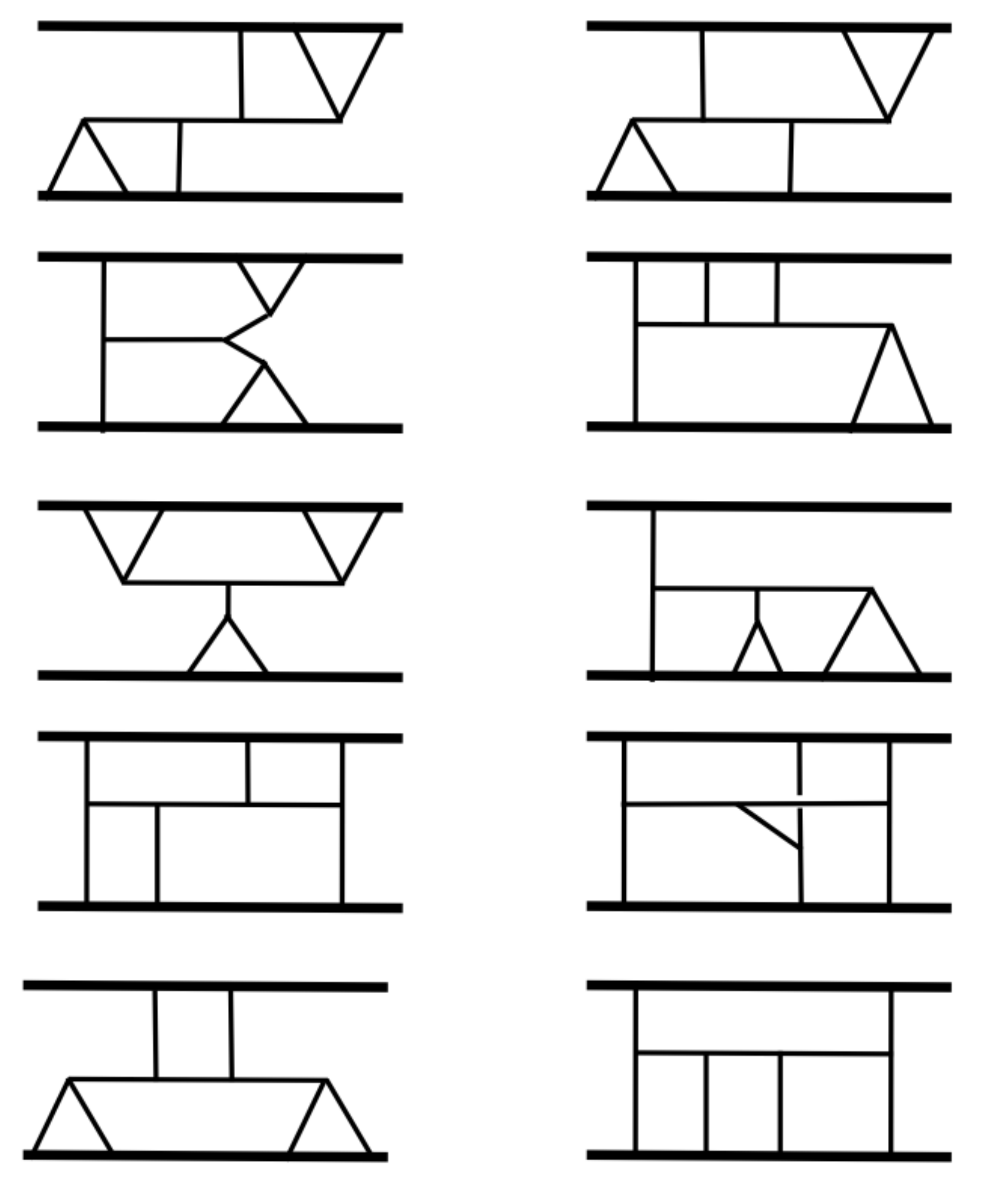}
\caption{The 25 irreducible $G_N^5$ topologies that contribute at 4PN. Each of them generate just one 4PN diagram. Other 25 4PN diagrams can be obtained from $G_N^5$ reducible topologies which have not been shown here.}
\label{topG5}
\end{figure}

At 4PN there are 515 new diagrams, variously distributed among the old topologies, new factorisable ones, as well as 12 new irreducible $G_N^4$ (see fig.~\ref{topG4}) topologies and 25 $G_N^5$ ones (fig.~\ref{topG5}).
Table \ref{tabdiatop} gives an overview of the topology and diagram counting. 
The corresponding Lagrangian has been computed  for the first time in \cite{Foffa:2012rn} up to terms of order $G_N^2$,
a sector which was subsequently also covered in the ADM framework \cite{Jaranowski:2012eb,Jaranowski:2013lca}.

\begin{table}
\begin{tabular}{l | c  c  c  c  c}
 & 0PN & 1PN & 2PN & 3PN & 4PN\\
\hline
$G$  & 1 &  &  &  &\\
$G^2$ &  & 1 & 1 & &\\
$G^3$ &  &  & 5 & 4 &\\
$G^4$ &  &  &  & 8 & 21\\ 
$G^5$ &  &  &  &  & 50\\
\end{tabular}
\quad\quad\quad\quad\quad\quad
\begin{tabular}{l | c  c  c  c}
 & $v^0$ & $v^2$ & $v^4$ & $v^6$\\
\hline
$G$  & 1 & 1 & 1 & \\
$G^2$ & 1 & 8 & 7 & 7\\
$G^3$ & 5 & 48 & 159 & \dots\\
$G^4$ & 8 & 299 &  \dots &\\
$G^4$ & 50 & \dots &  &\\
\end{tabular}
\caption{On the left: number of topologies entering at a given PN order. 
On the right: number of diagrams that start to contribute to the effective 
action at a given $G_N$ and $v$ power, for a total of 595 diagrams up to 4PN. From \cite{Foffa:2011ub}.}\label{tabdiatop}
\end{table}
Coming to step (iii), we have to perform perturbatively the functional 
integration of eq.~(\ref{eq:Feyn}).
As an illustration we take the contribution given by the second diagram of 
fig.~\ref{diaG2PN2}. The exponential in the functional integral has to be
expanded to the fourth order and the four Lagrangian terms corresponding to the
vertices present in the diagram have to be selected:
\be
\label{genamptime}
\ba{rcl}
\ds i S_{eff}&\supset&\ds -i V_{ex}\equiv \log\int {\cal D}\phi{\cal D} A
e^{iS_{bulk-free}(\phi,A)}\\
&&\ds\frac 12\int_{t_{1a},t_{1b},t_{2},t,\vec{x}}
\frac{-i m_1 \phi_{1a}}{\Lambda}\frac{-i m_1 \phi_{1b}}{\Lambda}
\frac{i m_2\vec{A}_2\cdot\vec{v_2}}{\Lambda}
\frac{-2ic_d\dot{\phi}\vec{A}\cdot\vec{\nabla}\phi}{\Lambda}\,,
\ea
\ee
where $S_{bulk-free}$ is the quadratic part of the bulk gravity action, 
$\phi\equiv\phi(t,{\bf x})$ and $\phi_{1a}\equiv\phi(t_{1a},\vec{x_1}(t_{1a}))$ and
so on.
Performing the Gaussian integral in the above eq.~(\ref{genamptime}) boils
down to substituting pair of like-fields with Green functions like in 
eq.~\ref{eq:effG} (indicated below with a contraction, as the procedure is in 
complete analogy to the Wick theorem in quantum field theory computations):
\be
-i V_{ex}=-\frac{m_1^2 m_2 c_d}{\Lambda^4}\int_{t_{1a},t_{1b},t_{2}}2v_2^i
\bcontraction{}{\phi_{1a}}{}{\dot{\phi}}
\bcontraction{\phi_{1a}\dot{\phi}}{\phi_{1b}}{}{\phi^{,j}}
\bcontraction{\phi_{1a} \dot{\phi} \phi_{1b} \phi^{,j}}{A_j}{}{A_{2; i}}
\phi_{1a} \dot{\phi} \phi_{1b} \phi^{,j} A_jA_{2; i}\,.
\ee
Expressing the Green functions in the momentum space via eqs.~(\ref{Fourk} - \ref{propexp}) one has
\be
-i V_{ex}&=&- \frac{m_1^2 m_2}{4 c_d \Lambda^4}\int_{t,t_{1a}}\int_{{\bf k}, {\bf k_1}}\frac{i\delta'(t-t_{1a}){\bf k_1}\cdot v_2(t)}{{\bf k}^2 ({\bf k}-{\bf k_1})^2 {\bf k_1}^2}
{\rm e}^{i\left[({\bf k}-{\bf k_1})\cdot x_1(t_{1a})+{\bf k_1}\cdot x_1(t)-{\bf k}\cdot x_2(t)\right]}\nonumber\\
&=&- \frac{m_1^2 m_2}{4 c_d \Lambda^4}\int_{t}\int_{{\bf k}, {\bf k_1}}\frac{i{\bf k_1}\cdot v_2(t)}{{\bf k}^2 ({\bf k}-{\bf k_1})^2 {\bf k_1}^2}i({\bf k}-{\bf k_1})\cdot v_1(t){\rm e}^{i{\bf k}\cdot r}\,.
\ee
A look at the structure of the denominator tells us that the complexity of this diagram (and of all the diagrams derived from the same topology) is equivalent to $1$-loop diagrams in quantum field theory (QFT);
indeed, this amplitude can be easily evaluated using standard textbook formulae
and taking the limit $d\rightarrow 3$
(in this case the amplitude is finite in dimensional regularization, so $d=3$ 
could have been set from the beginning), thus bringing to the following term of the 2PN action
\be
V_{ex}=-\int_t \frac{G_N^2 m_1^2 m_2}{r^2}\left[v_1^r v_2^r-v_1.v_2\right]\,,\quad v_i^r\equiv \frac{r.v_i}{r}\,.
\ee
Naturally, the same diagram contributes also to higher PN's, and the corresponding amplitude
is obtained as above with the caution of including the appropriate orders in in the $v$ expansion of $S_{pp}$ from eq.~(\ref{matter_grav}) and in the propagators expansions, eq.~(\ref{propexp}).
The latter may generally bring more and more ${\bf k},{\bf k_1}$ terms in the integrand numerator thus making the evaluation more lengthy, but as the general structure of the denominator does not change,
the complexity of momentum integrals remains still comparable to $1$-loop ones.

All amplitudes can be expressed in terms of (eventually complicated) spatial momentum integrals along the same lines. The actual evaluation strategy of the
integrals depends on the topology and, as we have seen, all the topologies up to $G_N^2$ can be computed by directly applying standard textbook formulae.
Generally, a $G_N^{(n+1)}$ irreducible topology is expected to involve momentum integrals equivalent to $n$-loops QFT diagrams, but
a more careful inspection shows that the situation is actually more favorable.
For instance, four of the five irreducible $G_N^3$ topologies in fig.~\ref{topG3} involve \emph{nested} loops integrations, that is integrals where at least one
${\bf k_a}$ appear just twice in the denominator: in this case this variable can be integrated out immediately as in the 1-loop case, and the result of the partial integration is,
in the $G_N^3$ case, easily integrable in terms of the remaining momentum variables. The only apparent exception to this rule is the H-shaped topology in fig.~\ref{topG3},
but an appropriate use of Integration by Parts techniques provide the following useful relation
\be\label{Hamp}
\!\!\!\!\!\!I(\alpha,\beta,\gamma,\delta,\epsilon)&\equiv&\int_{{\bf k_1},{\bf k_2}}\left[{\bf k_1}^{2\alpha}({\bf k}- {\bf k_1})^{2\beta}{\bf k_2}^{2\gamma}({\bf k}- {\bf k_2})^{2\delta}({\bf k_1}-{\bf k_2})^{2\epsilon}\right]^{-1}\nonumber\\
&=&\frac{\gamma\left[I(\alpha-,\gamma+)-I(\epsilon-,\gamma+)\right]+\delta\left[I(\beta-,\delta+)-I(\epsilon-,\delta+)\right]}{2\epsilon+\gamma+\delta-d}\,,
\ee
with the notation $I(\alpha-,\gamma+)\equiv I(\alpha-1,\beta,\gamma+1,\delta,\epsilon)$,
by means of which the integrals of this topology can be reduced to nested loops ones.
Thus, the $G_N^3$ sector does not present new conceptual difficulties with respect to the $G_N^2$ one,
although the computational challenge
becomes relevant at high PN because of the high number of diagrams involved, see tab.~\ref{tabdiatop},
and of the appearance of more and more ${\bf k_a}$ factors in the numerators.

The situation is somehow similar in the $G_N^4$ case, as it turns out that the topologies of this order involve, in the most difficult case, $3$-loops integrals which are
either nested or reducible through integrations by parts to integrals like the one in eq.~(\ref{Hamp}). 
Consequently one has the remarkable result that
all the topologies up to $G_N^4$ are basically tractable in terms of $1$-loop equivalent QFT diagrams, see also
\cite{Kol:2013ega} for related work.

At $G_N^5$ however things change, for two reasons: first, the use of integration by parts becomes more complicated and 
substantially intractable by hand.
This problem can be overcome by using automated reduction packages which are routinely used in particle physics multi-loop 
calculations, see e.g.~\cite{vonManteuffel:2012np}.
Second, and more important, the ``miracle'' according to which everything could be ultimately reduced to $1$-loop integrals
does not take place anymore: in the worst cases,
that is for the topologies in $(row,column)=(3,2)$ and $(4,5)$ in fig.~\ref{topG5}, one is left even after integration by parts
with integrals equivalent to a $4$-loop mass-less QFT diagram, which has to be
evaluated in $d\sim 3$ by means  of ${\it ad-hoc}$ techniques \cite{FMSS}.
A possibly more efficient way to reorganize the diagrams have been proposed in \cite{Kol:2009mj},
while a radically different computational method has been recently suggested in \cite{Neill:2013wsa}.

Starting from 3PN, divergences appear in the form of $(d-3)$ poles:
\be
{\cal L}^{3PN}_{pole}=-\frac{11G_N^2m_1^2m_2}{2(d-3)}\left[a_1^2+2a_1.a_2\right]+\frac{11G_N^3m_1^2m_2^2}{3(d-3)}a_1^r+\left(1\leftrightarrow 2\right)\,.
\ee
This divergence is not due to a short-distance incompleteness of the effective
field theory approach, and it has been found in all the past treatments at 3PN
with different kind of regularisations, see 
\cite{Damour:2001bu,Itoh:2003fz,Blanchet:2000nv,Blanchet:2000ub,deAndrade:2000gf}.
Since a Lagrangian is not an observable we can allow divergent terms in it
as long as any relation among observables is given by finite expressions: e.g. 
this singularity does not appear in the expression for $E(\omega)$ relating the 
energy $E$ of the system to the orbital angular velocity $\omega$.
It is however more practical to deal with a finite quantity also at the
Lagrangian level and this can be obtained at 3PN by means of the following 
word-line re-parametrization:
\be
\vec{x}_{1,2}\rightarrow\vec{x}_{1,2}+\frac{G_N^2m_{1,2}^2}{3}\vec{a}_{1,2}\,,
\ee
see \cite{lrr-2006-4} and  \cite{Foffa:2011ub} for details.

The EFT approach allowed us to compute for the first time the dynamics at 4PN up to ${\cal O}(G_N^2)$ (while some sectors at higher $G_N$ order have been recently covered in the ADM framework \cite{Jaranowski:2013lca}).
We write here the expression of the energy in the center of mass frame, addressing to \cite{Foffa:2012rn} for other details:
\be
\label{eq:4PN}
E^{4PN}&=& \frac{9\mu v^{10}}{256}\left(7-121 \nu+785 \nu^2-2254 \nu^3+2415 \nu^4\right)\nonumber\\
&&+\frac{G M \mu}{128 r}\left[v^8\left(525-4011 \nu+9507 \nu^2-714 \nu^3-15827 \nu^4\right)\right.\nonumber\\
&&-4v^6 {v^r}^2\left(147-369 \nu-1692 \nu^2+4655 \nu^3\right)\nu\nonumber\\
&&+18v^4 {v^r}^4\left(3+54 \nu-374 \nu^2+539 \nu^3\right)\nu\nonumber\\
&&+20 v^2{v^r}^6\left(5-50 \nu+148 \nu^2-119\nu^3\right)\nu\nonumber\\
&&\left.-35{v^r}^8\left(1-7 \nu+14 \nu^2-7\nu^3\right)\nu\right]\nonumber\\
&&+\frac{G^2 M^2 \mu}{1920 r^2}\left[15v^6\left(2300-4489 \nu+10258 \nu^2-16478 \nu^3-7800\nu^4\right)\right.\nonumber\\
&&+15v^4 {v^r}^2\left(120-5983 \nu-25990 \nu^2+37022 \nu^3+22760\nu^4\right)\nonumber\\
&&+5v^2 {v^r}^4\left(5347+77860 \nu-21072 \nu^2-25920 \nu^3\right)\nu\nonumber\\
&&\left.-3{v^r}^6\left(4771+36880 \nu+5440 \nu^2-4800\nu^3\right)\nu\right]+{\cal O}(G_N^3)\,,
\ee
with the symmetric mass ratio given by $\nu\equiv m_1m_2/M^2$.

Specializing then to circular orbits, that allows 
to express both $v$ and $G_NM/r$ in terms of $x\equiv(G_NM\omega)^{2/3}$, at
4PN one has
\be
\label{eq:Ec4PN}
\!\!\!\!\!\!\!\!\!\!\!\!\!\!\!\!\! E(x)|_{4PN}&=&-\mu \frac{x^5}2\left[-\frac{3969}{128}+\pa{\frac{448}{15}\log(x)
-\frac{123671}{5760}+\frac{9037}{1536}\pi^2+\frac{1792}{15}\log 2+
\frac{896}{15}\gamma}\nu+\right.\nonumber\\
&+&\left.\pa{-\frac{498449}{3456}+\frac{3157}{576}\pi^2}\nu^2+\frac{301}{1728}\nu^3+\frac{77}{31104}\nu^4\right]\,,
\ee
where $\gamma\simeq 0.577\ldots$ is the Eulero-Mascheroni constant.
Eq.~(\ref{eq:4PN}), together with inputs from Lorentz invariance of the 3PN
Lagrangian, allows to derive the $\nu^3$ and $\nu^4$ term in the above 
eq.~(\ref{eq:Ec4PN}), first obtained in \cite{Jaranowski:2012eb}, while the 
$\nu^2$ term has been obtained more recently in 
\cite{Jaranowski:2013lca}. The term linear
in $\nu$ has been obtained within the extreme mass ratio limit approach in 
\cite{Blanchet:2010zd,LeTiec:2011ab}, and its non-logarithmic part has been analytically
computed in \cite{Bini:2013zaa}. We shall discuss in subsec.~\ref{sse:radreac}, 
how the logarithmic piece can be derived from radiation reaction computation.
The $\nu$-independent part can be derived from the Schwarzschild result.

\subsection{Spin}
\label{sse:conspin}
EFT methods are giving a relevant contribution to the study of the spin sector of compact binary systems:
the next-to leading order (NLO) dynamics with a quadratic dependence on the stars' spins has been computed for
the first time in \cite{Porto:2006bt,Porto:2007px,Porto:2008jj},
triggering a renewed attention on such sector and a healthy competition with more traditional approaches, which led to the confirmation of the new results and even to the extension to NNLO
for the $S_1S_2$ potential \cite{Hartung:2011ea,Levi:2011eq} and for spin-orbit
\cite{Hartung:2011ea,Marsat:2012fn,Bohe:2012mr,Bohe:2013cla}.

As the spin of a compact star and the lowest-order spin-orbit and spin-spin interactions scale respectively like
\be
S\sim M v_{\rm rot} R_s\,,\quad V_{SO}\sim \frac{G_N M}{r^2} v.S\,,\quad V_{S^2}\sim \frac{G_N}{r^3}S_1.S_2\,,
\ee
one deduces that the lowest order (LO) spin orbit potential is a 1.5PN term for maximally rotating objects ($v_{\rm rot}\sim 1$), while the LO spin-spin interaction starts at 2PN.
 
Spin interactions in general relativity are introduced by means of the tetrad $e^\mu_a$ (for a more detailed discussion, see the papers cited in this section and \cite{Porto:2005ac,Porto:2008tb,Porto:2010tr}).
which transforms the metric into a locally free-falling (and locally Lorenz-invariant) frame:
\be
g_{\mu\nu}e^\mu_a e^\nu_b=\eta_{ab}\,.
\ee
If such frame is also chosen to be co-rotating with the spinning body, the tetrad geodesic variation is locally a rotation with generalized angular velocity given by
\be
\frac{{\rm d}e^{a\mu}}{{\rm d}\tau}\equiv u^\rho e^{a\mu}_{;\rho}=\Omega_{\nu}^{\mu} e^{a\nu}\Longrightarrow\Omega^{\mu\nu}=
e^\mu_a\frac{{\rm d}e^{a\nu}}{{\rm d}\tau}=-\Omega^{\nu\mu}\,,
\ee
where $u^\rho$ is the four velocity of the spinning body.
Local coordinate, Lorentz and parametrization invariances require the Lagrangian to be made of invariant contractions of $\Omega^{\mu\nu}$, $u^\rho$ and eventually of the local curvature tensors,
but do not unambiguously fix its form even in the case of flat space-time.
However it turns out that if one neglects finite-size effects,
the variation of any possible Lagrangians w.r.t. to the spinning body local position and tetrad, when expressed in terms of the conjugate momenta
$p^\mu=\frac{\delta{\cal L}}{\delta u_\mu}$ and $S^{\mu\nu}=\frac{\delta{\cal L}}{\delta \Omega_{\mu\nu}}$, gives the same (Mathisson-Papapetrou) equations of motion:
\renewcommand{\arraystretch}{1.6}
\be
\ba{rcl}
\ds\frac{{\rm d}p^\mu}{{\rm d}\tau}&=&\ds -\frac{1}{2}R_{\mu\nu\rho\sigma}u^\nu S^{\rho\sigma}\,,\\
\ds\frac{{\rm d}S^{\mu\nu}}{{\rm d}\tau}&=&\ds p^\mu u^\nu-p^\nu u^\mu\,.
\ea
\ee
\renewcommand{\arraystretch}{1.}

Since the spin is related to the conjugate momentum $S^{\mu\nu}$ rather than to the fundamental tetrad variables themselves, it is actually more convenient to work with a 
functional that behaves as an Hamiltonian with respect to the spin, while remaining a Lagrangian with respect to the body position $x^\mu$.
Such functional is called a Routhian and one can verify that the following form involving the spin connection $\omega_{\mu}^{ab}\equiv e^{b\nu} e^a_{\nu;\mu}$
\be\label{routhian}
{\cal R}_0=-m\sqrt{-u^2}-\frac{1}{2}S_{ab}\,\omega_{\mu}^{ab}u^\mu\,,
\ee
gives exactly the Mathisson-Papapetrou equations
by means of
\be
\frac{\delta}{\delta x^\mu}\int{\rm d}t R=0\,,\quad \frac{{\rm d}S^{ab}}{{\rm d}\tau}=\left\{{\cal R},S^{ab}\right\}\,,
\ee
once the following Poisson bracket is taken into account:
\be
\left\{S^{ab},S^{cd}\right\}=\eta^{ac}S^{bd}+\eta^{bd}S^{ac}-\eta^{ab}S^{cd}-\eta^{cd}S^{ab}\,.
\ee

The antisymmetric tensor $S^{\mu\nu}$ (which appears above through its locally flat-frame components $S^{ab}\equiv S^{\mu\nu}e_\mu^a e_\nu^b$) is the generalized spin of the body and it contains redundant degrees of freedom.
The redundancy corresponds to the ambiguity related the choice of a reference
world-line inside the body. One can reduce from 6 to the 3 degrees of freedom needed to describe 
an ordinary spin vector by imposing the Spin Supplementary Condition (SSC), which
relates the vector $S^{0i}$ to the physical spin components $S^{i}\equiv\varepsilon^{ijk} S_{jk}$.
There is not a unique way to impose such condition and the so-called covariant SSC
\be\label{covSSC}
S^{\mu\nu}p_\nu=0
\ee
will be taken here. The requirement of SSC conservation along the word line
gives the following relation:
\be\label{SSCons}
p^\mu=m \frac{u^\mu}{\sqrt{-u^2}}+\frac{1}{2m}R_{\nu\beta\rho\sigma}S^{\mu\nu}S^{\rho\sigma}\frac{u^\beta}{\sqrt{-u^2}}
+{\cal O}(R_{\nu\beta\rho\sigma}^2)\,,
\ee
where the first term in the r.h.s.~is SSC-independent and gives the familiar dynamics for a non spinning body.

Such relation can be enforced at the level of the Routhian by adding
\be
{\cal R}_{SSC}\equiv-\frac{1}{2 m}R_{abcd}S^{cd}S^{ae}\frac{u^{b}u^e}{\sqrt{-u^2}}+{\cal O}(R_{\nu\beta\rho\sigma}^2)
\ee
at the l.h.s. of eq.~(\ref{routhian}).

It should be remarked that imposition of the SSC implies $S^{0i}\sim S^{ij}v_j$ thus providing different scalings
for the different components of the spin tensor.
Being an algebraic constraint, the SSC can be imposed by direct replacement of $S^{0i}$ indifferently at the level of the fundamental Routhian or in the effective potential or in the equations of motion:
the second option will be followed here because it simplifies intermediate calculations, at the price however of some loss of transparency in the results,
which will not have a transparent physical interpretation until the SSC will be enforced.

Spin-induced finite size-effects become relevant much before than in the spin-less case;
the lowest order of these effects is the spin-induced quadrupole moment,
which can be taken into account by the following Routhian term
\be
\label{eq:rfs}
{\cal R}_{fs}\equiv\frac{C_{ES^2}}{2 m}\frac{E_{ab}}{\sqrt{-u^2}}S^a_c S^{cb}\,,
\ee
where $E_{ab}$ is the electric part of the Weyl tensor, and $C_{ES^2}=1$ for black
holes, while it has to be fixed via a matching procedure in the non-BH case.
This term gives an effective contribution to the $I_{ij}E_{ij}$ interaction in 
eq.~(\ref{eq:ext}) already at 2PN order.

The spin-dependent part of the Routhian can be expressed as follows in terms of the Kaluza-Klein fields:
\be\label{RouNRG}
\!\!\!\!\!\!\!\!\!\!\!\!\!\!\!\!\! {\cal R}&\supset&S^{ij}\left\{\frac{1}{4}F_{ij}\left(1+4\phi+8\phi^2\right)+\frac{1}{2}A_j \phi_{,i}\left[1+3\phi\right]+\frac{1}{2}A_j v_i \dot{\phi}
+\frac{1}{4}F_{jk}\sigma_i^k+\frac{1}{8}A_i \dot{A}_j\right.\nonumber\\
&+&\left.\left[A_j A_{i,k}-2F_{ij}A_k+2A_i A_{k,j}\right]\frac{v^k}{8}+\phi_{,j}v_i+\frac{1}{2}\sigma_{jk,i}v^k+\frac{1}{2}\sigma_{ik}\left(\phi_{,j}v^k+\phi^{,k}v_j\right)\right\}\nonumber\\
&+&S^{0i}\left\{\frac{1}{2}\dot{A}_i\left(1+3\phi\right)+\left[\phi_{,i}-\frac{1}{4}F_{ij}v^j\right]\left(1+2\phi\right)+\frac{1}{2}\dot{\sigma}_{ij}v^j+\frac{1}{4}F_{ij}A^j-\frac{3}{2}A_k v^k \phi_{,i}\right.\nonumber\\
&+&\left.\frac{1}{2}A^k\phi_{,k}v_i+\frac{1}{2}\left(\phi A_{i,k}-A_i\phi_{,k}\right)v^k+\frac{1}{2}A_i \dot{\phi}-\frac{1}{2}\sigma_{ij}\phi_{,j}\right\}+\frac{1}{2 m}S^{jk}S^{il}A_{k,ij}v_l\nonumber\\
&+&\frac{C_{ES^2}}{2}\left\{\left[\left(\vec{\nabla}\phi\right)^2+\vec{a}\!\cdot\!\vec{\nabla}\phi\right]\left(S^{kl}\right)^2+2S^{0k}S^{jk}\phi_{,ij}v^i+S^{0i}S^{0j}\phi_{,ij}\right.\nonumber\\
&+&\left.\left[\phi_{,ij}(1+2\phi)+2\phi_{,i}\phi_{,j}+A_{l,ij}v^l+2\phi_{,i}v_j+\frac{3}{2}\phi_{,ij}v^2+2\phi_{,il}v_jv^l+\frac{1}{4}F_{ik}F_{jk}\right]S^{ik}S^{kj}\right\}\,,\nonumber\\
\ee
where $d$ has been set to $3$ as all the results obtained so far from this Routhian are at most next-to-next-to leading order and thus finite.
By analogy to the  spin-less case, (gauge-dependent) divergences are expected 
to appear at next-to-next-to-next-to leading order, corresponding to 4.5PN for 
spin orbit, and to 5PN for spin-quadratic interactions.

The determination of the effective potential proceeds along the same lines of the spin-less case, with the new Feynman rules dictated by (\ref{RouNRG}).
Spin insertions in the diagrams introduce PN penalty factors, making 
the integrals to be computed easier than the ones without spin at the same PN 
order, while
the physical interpretation of the results is made less transparent in the
spinning case.

\begin{figure}
\includegraphics[width=1\linewidth]{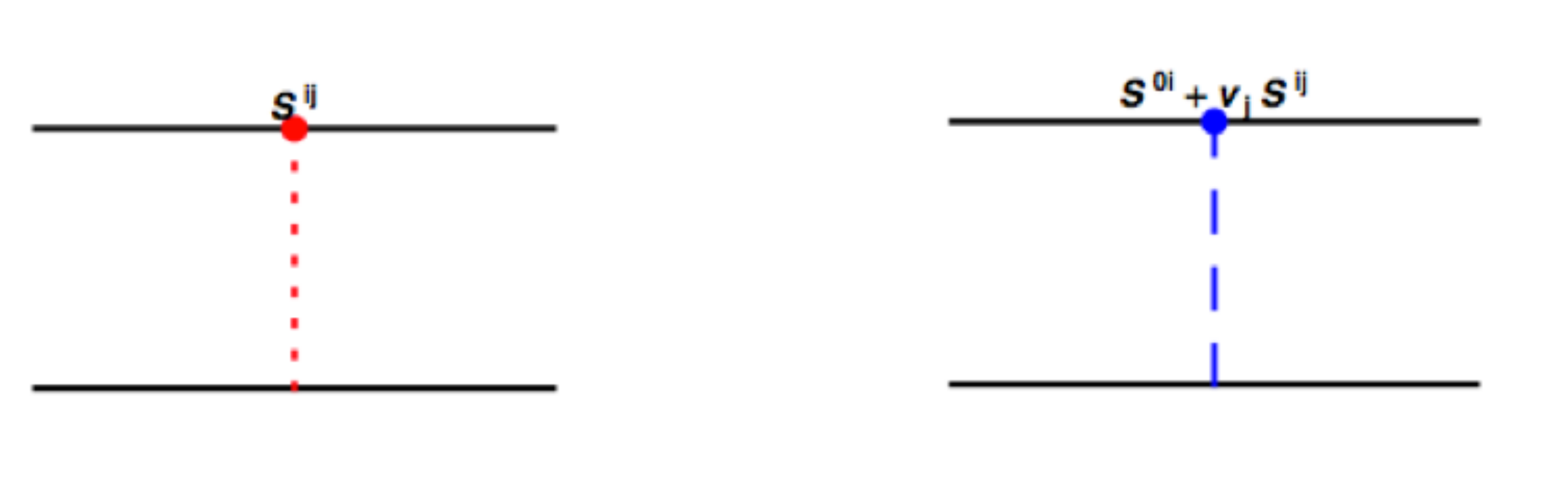}
\caption{Graphs contributing at leading order to the Spin-Orbit potential.}
\label{LOSpinOrbit}
\end{figure}

To illustrate the latter point, let us consider the lowest order spin-orbit 
interaction. According to the scaling rules (and reminding that 
$S^{0i}\sim v_jS^{ij}$), the effective potential is a 1.5 PN contribution that
can be derived from the two graphs in fig.~\ref{LOSpinOrbit} and their mirror 
images.
The computation is straightforward and gives
\be
V^{SO}_{LO}=-\frac{G_N m_2}{r^3}\left[\vec{S}_1\cdot(\vec{v}_1-2\vec{v}_2)\wedge \vec{r}+S^{0i}_1 r_i\right]+(1\leftrightarrow 2)\,.
\ee
The non-physical degrees of freedom represented by $S^{0i}$ must now be
eliminated through a SSC, as for instance the covariant one in 
eq.~(\ref{covSSC}).
By taking such condition at leading order in $v$ one gets
\be
V^{SO}_{LO}=-2\frac{G_N m_2}{r^3}\vec{S}_1\cdot\vec{v}\wedge \vec{r}+(1\leftrightarrow 2)\,,
\ee
which however does not correspond to the canonical result, see e.g. \cite{Damour:2007nc}:
\be\label{SOcan}
V^{SO}_{LO}=2\frac{G_N m_2}{r^3}\vec{S}_1\cdot\vec{v}\wedge \vec{r}+\frac{1}{2}\vec{S}_1\cdot\vec{v}_1\wedge \vec{a}_1+(1\leftrightarrow 2)\,.
\ee
The mismatch does not lead to any difference in physical observables, as it can be cured by means of the following spin-dependent coordinate transformation at the Lagrangian level
\be
\vec{x}_{1,2}\rightarrow \vec{x}_{1,2}+\frac{1}{2 m_{1,2}}\vec{S}_{1,2}\wedge \vec{v}_{1,2}\,.
\ee
Alternatively, an expression matching exactly eq. (\ref{SOcan}) may be obtained by imposing the so-called Newton-Wigner SSC, $S^{\mu\nu}\left(p_\nu+m e^0_\nu\right)=0$, as well as the Newtonian equations of motion for the accelerations \cite{Barausse:2009aa}.

To summarize, the choice of working with $S^{0i}$ at the effective potential level makes the results formally SSC-dependent, but the difference vanishes on observables.
Clearly when going at higher PN orders one should not forget to include effects coming from the higher order terms in the SSC relation eventually inherited from lower PNs.
An alternative procedure is to impose the SSC directly at the level of the fundamental Routhian, a strategy which however brings unnecessary complications in the
intermediate steps of the calculation.
Whatever choice is made, one is left with some difficulties in comparing results
derived within different approaches, like the EFT method and the ADM approach
(a problem somehow addressed for instance in \cite{Hergt:2012zx}).
Not surprisingly, such difficulties become computationally more relevant at 
higher post-Newtonian order, as is the case for the $S_1S_2$ 4PN sector, where 
a full comparison between the two approaches has not yet been carried on.

\section{Radiation}

\label{se:radiation}
In the previous section we have shown how to obtain an effective action \`a la 
Fokker describing the dynamics of a binary system at the orbital
scale $r$ in which gravitational degrees of freedom have been integrated out,
resulting in a series expansion in $v^2$, as in a conservative system
odd powers of $v$ are forbidden by invariance under time reversal.

The gravitational tensor in 3+1 dimensions has 6 physical degrees of freedom 
(10 
independent entries of the symmetric rank 2 tensor in 3+1 dimensions minus 4 
gauge choices): 4 of them are actually constrained, non radiative 
physical degrees of freedom, responsible for the gravitational potential, and the remaining 2 
are radiative, or GWs.

In order to compute interesting observables, like the average energy flux
emitted by or the radiation reaction on the binary system, it will be 
useful to ``integrate out'' also the radiative degrees of freedom, with
characteristic length scale $\lambda=r/v$, as it will be shown in the 
next subsections.

We aim now at writing the coupling of an extended source appearing in 
eq.~(\ref{eq:ext}) in terms of the energy momentum
tensor $T^{\mu\nu}(t,x)$ moments.
Here we use $T^{\mu\nu}$, as in \cite{Goldberger:2009qd}, to denote 
the term relating the effective action $\az_{1g}$ relative to the single graviton
emission
\be
\az_{1g}\propto\int {\rm d}t{\rm d}^dx\, T^{\mu\nu}(t,x)h_{\mu\nu}(t,x)\,,
\ee
to the gravitational mode generically denoted by $h_{\mu\nu}$.
With this definition $T^{\mu\nu}$ receives contribution from both matter
and the gravity \emph{pseudo-tensor} appearing in the traditional GR description
of the emisson processes.
 
Given that the variation scale of the energy momentum tensor and of the 
radiation field are respectively $r_{source}$ and $\lambda$,
by Taylor-expanding the standard term $T_{\mu\nu}h^{\mu\nu}$
\be
\label{eq:mexp}
\left.\sum_{n}\frac 1{n!}\dpa_1\ldots \dpa_n h_{\mu\nu}(t,x)\right|_{x=0}
\int {\rm d}^dx'\, T^{\mu\nu}(t,x')x'_1\ldots x'_n\,,
\ee
we obtain a series in $r_{source}/\lambda$, which for binary systems gives
$r_{source}=r\ll\lambda=r/v$.

The results of the integral in eq.~(\ref{eq:mexp}) are source moments that,
following standard procedures not exclusive of the effective field theory
approach described here, are traded for mass and velocity multipoles.
For instance, the integrated moment of the energy momentum tensor can be 
traded for the mass quadrupole
\be
Q_{ij}(t)\equiv\int {\rm d}^dx\, T_{00}(t,x)x_ix_j\,,
\ee
by repeatedly using the equations of motion under the form 
$T_{\mu\nu}^{\ \ ,\nu}=0$:
\be
\label{eq:T0i}
\ba{rcl}
\ds\int {\rm d}^dx\,\paq{T_{0i}x_j+T_{0j}x_i}&=&\ds\int  {\rm d}^dx\, T_{0k}\pa{x_ix_j}_{,k}\\
&=&\ds -\int {\rm d}^dx\, T_{0k,k}\,x_ix_j\\
&=&\ds \int {\rm d}^dx\, \dot T_{00}\,x_ix_j=\dot Q_{ij}
\ea
\ee
\be
\label{eq:Tij}
\ba{rcl}
\ds 2\int {\rm d}^dx\,T_{ij}&=&
\ds\int  {\rm d}^dx\,\paq{T_{ik}\,x_{j,k}+T_{kj}\,x_{i,k}}\\
&=&\ds\int  {\rm d}^dx\,\paq{\dot T_{0i}x_j+\dot T_{0j}x_i}\\
&=&\ds\int  {\rm d}^dx\,\ddot T_{00}x_ix_j=\ddot Q_{ij}\,.
\ea
\ee
The above equations also show that as for a composite binary system 
$T_{00}\sim O(v^0)$, then $T_{0i}\sim O(v^1)$ and $T_{ij}\sim O(v^2)$.

Taking as the source of GWs the composite binary system, the multipole series 
is an expansion in terms of $r/\lambda=v$, so when expressing the multipoles
in terms of the parameter of the individual binary constituents, powers of $v$ 
have to be tracked in order to arrange a consistent expansion.
At lowest order in the multipole expansion and at $v^0$ order
\be
\label{eq:Mon}
\az_{ext}|_{v^0}= \frac 1\Lambda\int  {\rm d}t\,{\rm d}^dx\,T_{00}|_{v^0}\,\phi=
\frac M\Lambda \int  {\rm d}t\,\phi\,,
\ee
where in the last passage the explicit expression
\be
T_{00}(t,x)|_{v^0}=\sum_A m_A\delta^{(3)}(x-x_A(t))\,,
\ee
has been inserted.
At order $v$ the contribution from the first order derivative in 
$\phi$ have to be added the contribution of $T_{\mu\nu}|_{v}$, which gives
\be
\az_{ext}|_{v}=\frac 1\Lambda \int  {\rm d}t\,{\rm d}^dx\,
\pa{T_{00}|_{v^0}\,x_i\phi_{,i}+T_{0i}|_{v}\,A_i}
\,,
\ee
with
\be
T_{0i}(t,x)|_{v}=\sum m_Av_{Ai}\delta^{(3)}(x-x_A(t))\,,
\ee
and neither $T_{00}$ nor $T_{ij}$ contain terms linear in $v$.
Since the total mass appearing in eq.~(\ref{eq:Mon}) is conserved (at this 
order) and given that in the center of mass frame 
$\sum_A m_Ax_{Ai}=0=\sum_A m_Av_{Ai}$, there is no radiation up to order $v$.
From order $v^2$ on, following a standard procedure, see e.g.~\cite{MaggioreGW},
it is useful to decompose the source coupling to the gravitational fields in 
irreducible representations of the $SO(3)$ rotation group, to obtain
\be
\label{eq:QE}
\ba{rcl}
\ds\left.S_{ext}\right|_{v^2}^\1&=&\ds-\frac 12\int  {\rm d}t\,{\rm d}^dx\,
T_{0i}|_{v}\,x_j(A_{i,j}-A_{j,i})\,,\\
\ds\left.S_{ext}\right|_{v^2}^{\0+\2}&=&\ds\frac 12\int  {\rm d}t\,Q_{ij}|_{v^0}
\pa{\ddot \sigma_{ij}-2\phi_{,ij}-\frac 2{d-2}\ddot\phi\,\delta_{ij}
-\dot A_{i,j}-\dot A_{j,i}}\,,
\ea
\ee
were eqs.~(\ref{eq:T0i},\ref{eq:Tij}) and integration by parts have been used,
$\0,\1,\2$ stand for the scalar, vector and symmetric-traceless representations 
of $SO(3)$, and
\be
\left.Q_{ij}\right|_{v^n}=\int  {\rm d}^dx\,T_{00}|_{v^n}\,x_ix_j\,.
\ee
The $\0+\2$ term in eq.~(\ref{eq:QE}) reproduces at linear order the $E_{ij}$ 
term in eq.~(\ref{eq:ext}), allowing to identify $I_{ij}$ with $Q_{ij}$ at 
leading order.

The $\1$ part matches the second term in eq.~(\ref{eq:ext}), and it is not 
responsible for radiation as it couples $A_i$ to the conserved angular momentum.
In order to simplify the calculation, we work from now on in the 
transverse-traceless (TT) gauge, in which the only relevant radiation field is 
the traceless and transverse part of $\sigma_{ij}$. The presence of the
other gravity polarizations is required by gauge invariance.

Discarding all fields but the TT-part of the $\sigma_{ij}$ field, at order $v^3$ 
one has
\be
S_{ext}|_{v^3}=\int  {\rm d}t\,{\rm d}^dx\,T_{ij}|_{v^2}\,x_k\,\sigma_{ij,k}
\ee
and using the decomposition \cite{MaggioreGW}
\be
\ba{cl}
\ds\int {\rm d}^dx\,T_{ij}x_k=&
\ds\frac 16\int  {\rm d}^dx\,\ddot T_{00}x^ix^jx^k\\
&\ds+\frac 13\int  {\rm d}^dx\, \pa{\dot T_{0i}x_jx_k+\dot T_{0j}x_ix_k-2\dot T_{0k}x_ix_j}\,,
\ea
\ee
we can re-write
\be
S_{ext}|_{v^3}=\int  {\rm d}t\pa{ \frac 16Q_{ijk}E_{ij,k}-\frac 23 P_{ij}B_{ij}}
\ee
where
\be
P_{ij}=\int  {\rm d}^dx\,\pa{\epsilon_{ikl}T^{0k}x^lx_j+\epsilon_{jkl}T^{0k}x^lx_i}\,,
\ee
and
\be
Q_{ijk}=\int  {\rm d}^dx\,T_{00}\,x_ix_jx_k\,,
\ee
allowing to identify $J_{ij}\leftrightarrow P_{ij}$ and 
$I_{ijk}\leftrightarrow Q_{ijk}$ at leading order.

At $v^4$ order the $T_{ij}x^kx^l\sigma_{ij,kl}$ term, beside giving the 
leading hexadecapole term (or $2^{\rm 4-th}$-pole), also gives a $v^2$ 
correction to the leading quadrupole interaction $I_{ij}E^{ij}$, which can be 
written as
\be
\!\!\!\!\!\!\!\!\!\!\!\!\!\!\!\!\!\!\!\!\!\!\!\!\!
S_{ext}|_{v^4}\supset\int  {\rm d}^{d+1}x\,\paq{T_{00}|_{v^2}+T_{kk}|_{v^2}-
\frac 43\dot T_{0k}|_{v}\,x^k+
\frac{11}{42}\ddot T_{00}|_{v^0}x^2}\pa{x^ix^j-\frac{\delta_{ij}}
dx^2}E_{ij}\,.
\ee
For the systematics at higher orders see \cite{Ross:2012fc} or the standard textbook 
\cite{MaggioreGW}.

\subsection{Matching between the radiation and the orbital scale}
In the previous subsection we have spelled out the general expression
of the effective multipole moments in terms of the energy-momentum tensor 
moments. However we have only used two ingredients from the specific binary
problem
\begin{itemize}
\item $T_{00}\sim m v^0$
\item the source size is $r$ and the length variation of the background is
$\lambda\sim r/v$\,.
\end{itemize}
Now we are going to match the coefficients appearing in eq.~(\ref{eq:ext}) with
the parameters of the specific theory at the orbital scale.

\begin{figure}
  \begin{minipage}[]{.49\linewidth}
    \begin{center}
      \includegraphics[width=\linewidth]{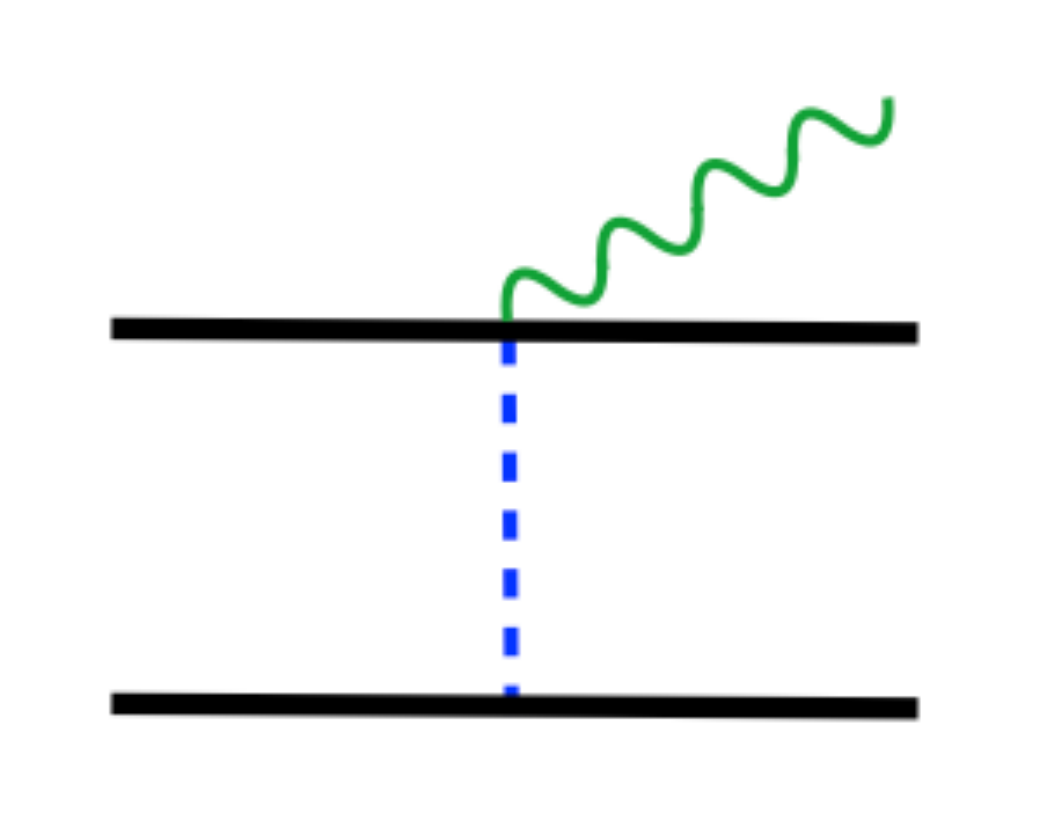}
    \end{center}
    \caption{Graph dressing $T_{00}$ at $v^2$ order.}
    \label{fi:t00v}
  \end{minipage}
  \begin{minipage}[]{.49\linewidth}
    \begin{center}
      \includegraphics[width=\linewidth]{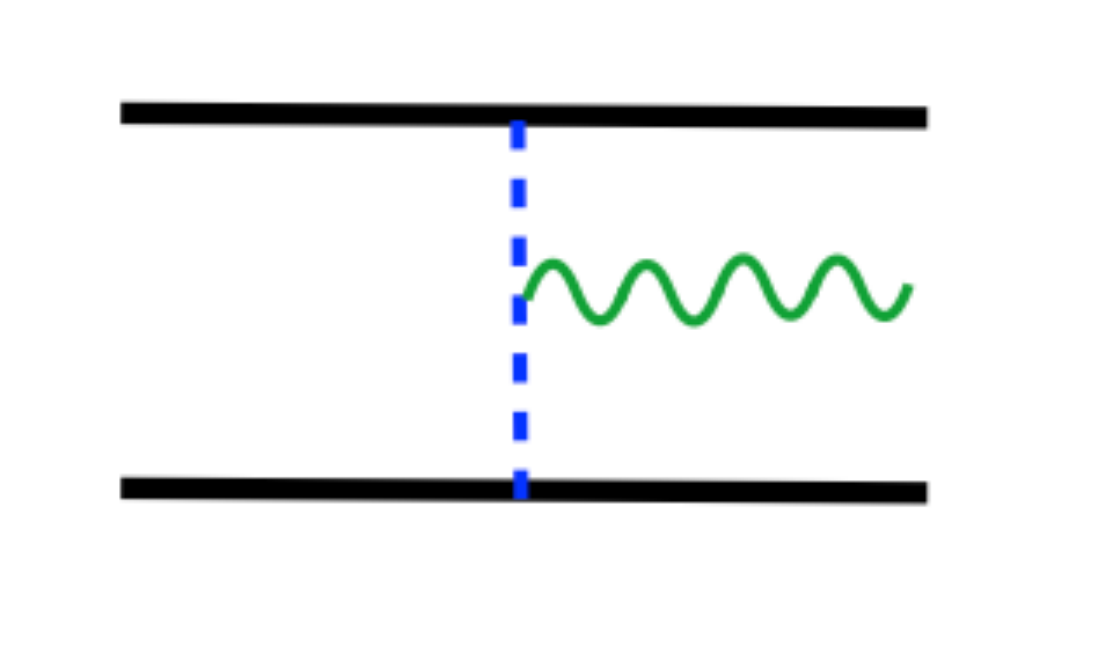}
    \end{center}
    \caption{Graph dressing $T_{ij}$ at leading order. The external radiation
    graviton does not carry momentum but it is Taylor expanded according
    to eq.~(\ref{eq:mexp}).}
    \label{fi:tijG}
  \end{minipage}
\end{figure}

At leading order $Q_{ij}|_{v^0}=\sum_A m_A x_{Ai}x_{Aj}$ and the $v^2$ corrections 
to $T_{00}$ can be read from diagrams in figs.~\ref{fi:t00v},\ref{fi:tijG}.
Such diagrams account for the pseudo-energy momentum tensor of the gravitational
field and are obtained by computing the effective action with the \emph{background
filed method}, and picking the term in the resulting effective action linerly coupled
to the background gravity field \cite{Goldberger:2004jt}.

As $\phi$ couples to $T_{00}+T_{kk}/(d-2)$ and $\sigma_{ij}$ to $T_{ij}$, from
the diagrams one obtains\footnote{Note that since only $\int T_{kk}$ is needed,
and not $T_{kk}$ itself, it could have been computed from eq.~(\ref{eq:Tij}) 
instead of from the diagram in fig.~\ref{fi:tijG}.}
\be
\ba{rcl}
\left.\ds\int {\rm d}^dx \pa{T_{00}+\frac 1{d-2}T_{kk}}\right|_{v^2}&=&\ds\sum_A
\frac d{2(d-2)}m_Av^2_A-g(d)\sum_{B\neq A}\frac{G_Nm_Am_B}{r^{d-2}}\,,\\
\ds\int d^{\rm d}xT_{kk}|_{v^2}&=&\ds\sum_A\frac 12m_Av_A^2-\frac{d-2}2g(d)
\sum_{B\neq A}\frac{G_Nm_Am_B}{r^{d-2}}\,,
\ea
\ee
where $g(d)= (d-2)\Gamma(d/2-1)/[\pi^{d/2-1}2^{d-4}(d-1)]$.
The calculation can be iterated for all higher multipoles, and it does not
contain any fundamental difference if framed within the effective field theory
approach or traditional methods.

\subsection{Spin contribution to the source moments}
In the case of spinning individual sources, in order
to add the spin contributions to the energy-momentum tensor we start from the 
spin-world-line term in eq.~(\ref{routhian}) to obtain
\be
\sqrt{-g}\,T^{\mu\nu}(t,x)=\frac 12\sum_A\dpa_\alpha\delta^{(3)}(x-x(t))
\pa{S_A^{\mu\alpha}u_A^\nu+S_A^{\nu\alpha}u_A^\mu}\,,
\ee
from which it is possible to derive \cite{Porto:2010zg} the leading order
energy momentum tensor components linear in the spins:
\be
\label{eq:ts2}
\ba{rcl}
\ds T^{00}(t,\K)|_{S^1}&=&\ds\sum_A S^{0i}_Ai\K^ie^{-i{\bf k}\cdot x_A}\,,\\
\ds T^{0i}(t,\K)|_{S^1}&=&\ds\frac 12\sum_A S^{ij}_Ai\K^j
e^{-i{\bf k}\cdot x_A}\,,\\
\ds T^{ij}(t,\K)|_{S^1}&=&\ds\frac 12\sum_A\pa{S^{il}_Av^j_A+S^{jl}_Av^i_A}i\K^l
e^{-i{\bf k}\cdot x_A}\,,
\ea
\ee
where a mixed coordinate-momentum space has been adopted, and the leading
$O(S_A^2)$ are given by
\be
T^{00}(t,\K)|_{S^2}=
-\sum_A\frac{C_{ES^2}^{(A)}}{2m_A}S^{ik}_AS^{jk}_A\K_i\K_je^{-i{\bf k}\cdot x_A}\,,
\ee
with $T^{ij}|_{S^2}\sim v T^{0i}|_{S^2}\sim v^2 T^{00}$.
Since $S^{0i}k\sim S^{ij}vk\sim mv^3$ (we recall that  $k\sim 1/r$ is
the wave number exchanged between binary constituents), the above components
of the energy momentum tensor can be used to compute the source moments
necessary to derive physical observables, as discussed in the next subsections.
At leading order in spin and $v$, the electric and magnetic quadrupole moments 
read (using the covariant SSC)
\be
\ba{rcl}
\ds I_{ij}|_{S^1}&\supset&\ds\sum_A\frac 83\epsilon^{ikl}\pa{v_{Ak}S_{l}x_{Aj}-
\frac 43x_{Ak}S_{l}v_{Aj}-\frac 43x_{Ak}\dot S_{Al}x_{Aj}+i\leftrightarrow j}\,,\\
\ds J_{ij}|_{S^1}&\supset&\ds\sum_A S_{Ai}x_{Aj}+S_{Aj}x_{Ai}\,,
\ea
\ee
where $S^i=\epsilon^{ijk}S_{kl}$.
For non-linear terms one has to add diagrams at the orbital scale analogous to 
figs.\ref{fi:t00v},\ref{fi:tijG} with spin insertion at the vertices, as well 
as the $O(S^2_A)$ term in the world-line energy momentum tensor in 
eq.~(\ref{eq:ts2}), which translates to quadratic terms in the quadrupole
moments given by
\be
\ba{rcl}
\ds I^{ij}|_{S_A^2}&\supset&\ds -\sum_A\frac{C_{ES^2}^{(A)}}{m_A}\pa{S^i_AS^j_A+S^j_AS^i_A}\,,\\
\ds J_{ij}|_{S_A^2}&\supset&\ds\sum_A \frac{C_{ES^2}^{(A)}}{m_A}
\pa{\epsilon_{ikl}v^kS^l_AS_{Aj}+i\leftrightarrow j}\,.
\ea
\ee

\subsection{Integrating out the radiating graviton: radiation reaction}
\label{sse:radreac}

\begin{figure}
  \begin{minipage}[htb]{.49\linewidth}
    \begin{center}
      \includegraphics[width=\linewidth]{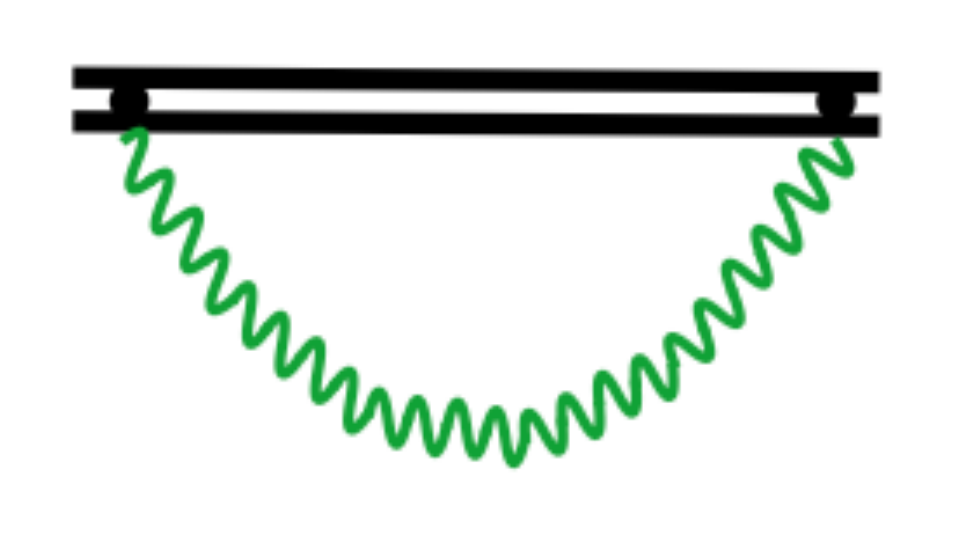}
    \end{center}
    \caption{Diagram giving the leading term of the amplitude describing radiation back-reaction on the sources.}
    \label{fi:quadRR}
  \end{minipage}
  \begin{minipage}{.49\linewidth}
    \begin{center}
      \includegraphics[width=\linewidth]{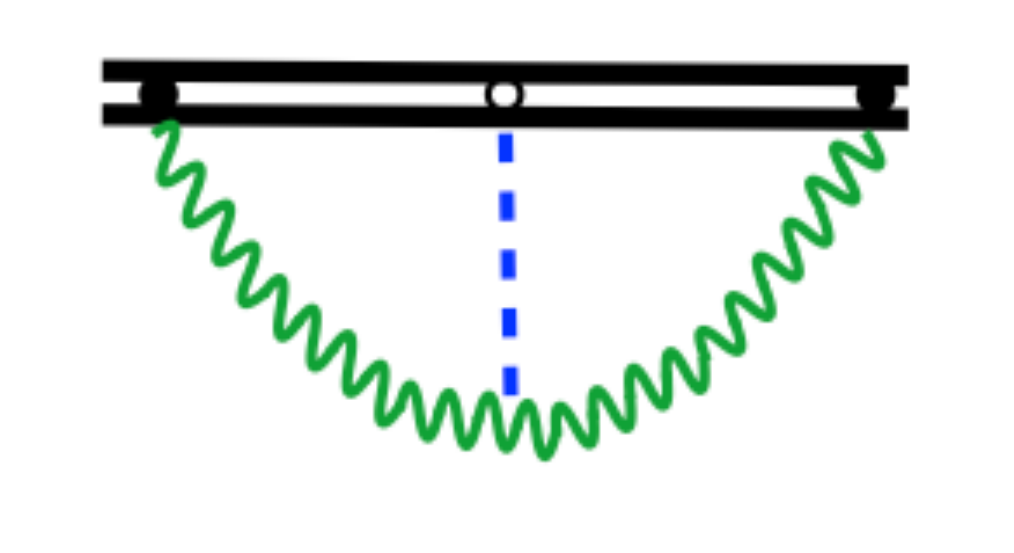}
      \caption{Next-to-leading order term in the back-reaction amplitude.}
      \label{fi:quadMRR}
    \end{center}
  \end{minipage}
\end{figure}

We have now built an effective theory for extended objects in terms of the 
source moments and also shown how to match the orbital scale with the theory
describing two point particles experiencing mutual gravitational attraction.
We can further use the extended object action in eq.~(\ref{eq:ext}) to integrate
out the gravitational radiation to obtain an effective action $S_{mult}$ for the
source multipoles alone.

In order to perform such computation, boundary conditions \emph{asymmetric in
time} have to be imposed, as no incoming radiation at past infinity is
required. Using the standard \emph{Feynman} propagator, which ensures a pure 
in-(out-)going wave at past (future) infinity, would lead to a non-causal 
evolution as it can be shown by looking at the following toy model 
\cite{Galley:2009px}, which is defined by a scalar field $\Psi$ coupled to
a source $J$:
\be
\label{eq:Aztoy}
S_{toy}=\int  {\rm d}^{d+1}x\,\paq{-\frac 12\pa{\dpa\psi}^2+\psi J}\,.
\ee
We may recover the field generated by the source $J$ as
\be
\label{eq:toy}
\psi(t,x)=\int  {\rm d}^{d+1}x\, G(t-t',x-x')J(t',x')\,,
\ee
where the Feynman propagator given by eq.~(\ref{eq:propT}) can also be written as
\be
G(t,x)=\theta(t)\Delta_+(t,x)+\theta(-t)\Delta_-(t,x)\,,
\ee
with $\Delta_\pm=e^{\mp i\omega t}e^{i{\bf k}\cdot x}/k$, which is clearly a-causal because
of the $\theta(-t)$ term. In a causal theory $\psi$ would be given by the same
eq.~(\ref{eq:toy}) but with the Feynman propagator replaced by the retarded one
$G_{Ret}(t,x)$, given by:
\be
\ba{rcl}
\ds G_{Ret}(t,x)&=&\ds-\int_{\bf k} \frac{{\rm d}\omega}{2\pi}
\frac{e^{-i\omega t+i{\bf k}\cdot x}}{{\bf k}^2-(\omega+i\epsilon)^2}\\
&=&\ds -i\theta(t)\paq{\Delta_+(t,x)-\Delta_-(t,x)}=G_{Adv}(-t,-x)\,.
\ea
\ee
However it is not possible to naively use the retarded propagator in the action
(\ref{eq:Aztoy}), as it would still yield non-causal equations of motions
\cite{Galley:2012hx}.
This problem was not present in the conservative dynamics described in 
sec.\ref{se:cons} as the Feynman Green function with symmetric boundary conditions is the
appropriate one to describe a conservative system.

However there is a consistent way to define an action for non-conservative 
system with asymmetric time boundary condition: by adopting a generalization of
the Hamilton's variational principle
similar to the closed-time-path, or in-in formalism (first
proposed in \cite{Schwinger:1960qe}, see \cite{deWitt} for a review) 
as described in \cite{Galley:2012hx}, which requires a \emph{doubling} of the 
field variables. For instance the toy model in eq.~(\ref{eq:Aztoy}) is modified 
so that
the generating functional for connected correlation functions in the in-in 
formalism has the path integral representation
\renewcommand{\arraystretch}{1.5}
\be
\label{eq:doubleP}
\!\!\!\!\!\!\! e^{i\mathcal{S}_{eff}[J_1,J_2]}&=&\ds\int \mathcal{D}\psi_1\mathcal{D}\psi_2 
\exp\pag{i\int {\rm d}^{d+1}x
\,\paq{-\frac 12(\dpa\psi_1)^2+\frac 12(\dpa\psi_2)^2-J_1\Psi_2+J_2\psi_2}}\,.\nonumber\\
\ee
\renewcommand{\arraystretch}{1.}
In this toy example the path integral can be performed exactly, and using the 
Keldysh representation \cite{Keldysh:1964ud} defined by
$\Psi_-\equiv\Psi_1-\Psi_2$, $\Psi_+\equiv (\Psi_1+\Psi_2)/2$,
one can write
\be
\mathcal{S}_{eff}[J_+,J_-]=\frac i2\int {\rm d}^{d+1}x\,d^{d+1}y
\,J_B(x)G^{BC}(x-y)J_C(y)\,,
\ee
where the $B,C$ indices take values $\{+,-\}$ and
\be
\label{eq:prop}
G^{BC}(t,\X)=
\pa{\ba{cc}
0 & iG_{Adv}(t,\X)\\
iG_{Ret}(t,\X) & \frac 12 G_H(t,\X)
\ea}\,,
\ee
where $G^{++}=0$ and $G_{Adv,Ret,H}$ are the usual advanced, retarded propagators
and Hadamard function respectively, with $G_H=\Delta_++\Delta_-$\,.
In our case, the lowest order expression of the quadrupole
in terms of the binary constituents world-lines $x_A$, i.e.
\be
\label{eq:Qlead}
Q_{ij}|_{v^0}=\sum_{A=1}^2 m_A\pa{x_{Ai}x_{Aj}-\frac{\delta_{ij}}d x_{Ak}x_{Ak}}\,,
\ee
is doubled to
\renewcommand{\arraystretch}{.6}
\be
\ba{rcl}
\ds Q_{-ij}|_{v^0}&=&\ds\sum_{A=1}^2m_A\pa{x_{-Ai}x_{+Aj}+x_{+Ai}x_{-Aj}}
-\frac 2d\delta_{ij}x_{+Ak}x_{-Ak}\,,\\
\ds Q_{+ij}|_{v^0}&=&\ds\sum_{A=1}^2m_Ax_{+Ai}x_{+Aj}-\frac 1d\delta_{ij}x_{+A}^2+O(x_-^2)\,.
\ea
\ee
\renewcommand{\arraystretch}{1.}
The word-line equations of motion that properly include radiation reaction 
effects are given by
\be
\label{eq:eqmoto}
\left.0=\frac{\delta S_{eff}[x_{1\pm},x_{2\pm}]}{\delta x_{A-}}\right|_{\stackrel{x_{A-}=0}{\small{x_{A+}=x_A}}}\,.
\ee

At lowest order, by integrating out the radiation graviton, i.e. by computing 
the diagram in fig.~\ref{fi:quadRR}, one obtains the Burke-Thorne 
\cite{BurkeThorne} potential term in the effective action $S_{mult}$
\be
\label{eq:BT}
S_{mult}|_{fig.~\ref{fi:quadRR}}=-\frac{G_N}5\int {\rm d}t\,Q_{-ij}(t)Q^{(5)}_{+ij}(t)\,,
\ee
where $A^{(n)}(t)\equiv {\rm d}^nA(t)/{\rm d}t^n$, which has been derived in the EFT 
framework in \cite{Galley:2009px}.
Corrections to the leading effect appears when considering as in the previous
subsection higher orders in the multipole expansion: the 1PN correction to the
Burke Thorne potential were originally computed in \cite{355730,Blanchet:1993ng}
and re-derived with effective field theory methods in \cite{Galley:2012qs}.

The genuinely non-linear effect, computed originally in 
\cite{Blanchet:1987wq,Blanchet:1993ng} and within effective field theory methods
in \cite{Foffa:2011np}, appears at relative 1.5PN order and it is due
to the diagram in fig.~\ref{fi:quadMRR}. The result turns out to have a 
short-distance singularity which introduces a 
logarithmic contribution to the effective action (by virtue of 
eq.~(\ref{eq:eqmoto}) only terms linear in $Q_-$ are kept)
\renewcommand{\arraystretch}{1.4}
\be
\label{eq:radReacRes}
\!\!\!\!\!\!\!
 S_{mult}|_{fig.\ref{fi:quadMRR}}&=&\ds -\frac 15G_N^2M\int_{-\infty}^\infty\frac{{\rm d}\omega}{2\pi}
\,\omega^6\ds\pa{\frac 1\epsilon-\frac{41}{30}+i\pi-\log\pi+\gamma+
\log(\omega^2/\mu^2)}\nonumber\\
&&\ds\paq{Q_{ij-}(\omega)Q_{ij+}(-\omega)+Q_{ij-}(-\omega)Q_{ij+}(\omega)}\,.
\ee
\renewcommand{\arraystretch}{1.}
We note the presence of the logarithmic term which is non-analytic in 
$k$-space and non-local (but causal) in direct space: after integrating
out a mass-less propagating degree of freedom the effective action is not
expected to be local \cite{Appelquist:1974tg}.
A local mass counter term $M_{ct}$ defined by
\be
M_{ct}=-\frac{2G_N^2}5M\pa{\frac 1\epsilon +\gamma -\log\pi}Q_{-ij}Q_{+ij}^{(6)} 
\ee
can be straightforwardly added to the world-line effective
action to get rid of the divergence appearing as $\epsilon\to 0$. 
According to the standard renormalization procedure, one can define a 
renormalized mass $M^{(R)}(t,\mu)$  for the monopole term in the action 
(\ref{eq:ext}), depending on time (or frequency) and on the arbitrary scale 
$\mu$ in such a way that physical quantities (like the energy or the radiation 
reaction force) will be $\mu$-independent\footnote{
Note that at the order required in the diagram in fig.\ref{fi:quadMRR}, 
$M^{(R)}(t,\mu)$ can be safely treated as a constant $M$ on both its arguments 
$t$ and $\mu$.}.

The derivation of the $\mu$ dependence of the renormalized mass was first 
obtained in 
\cite{Goldberger:2012kf} by evaluating $Q_{ij}^2$ corrections to the energy 
momentum tensor of a binary system.
Here we give a simplified version of such derivation, following 
\cite{Foffa:2011np}, by deriving the logarithmic corrections to the
equations of motion from eq.~(\ref{eq:radReacRes})
\be
\label{eq:resreg}
\left.\delta\ddot x_{Ai}(t)\right|_{log}=
-\frac 85 x_{aj}(t)G_N^2 M\int^t_{-\infty}{\rm d}t'\,Q_{ij}^{(7)}(t')\log\paq{(t-t')\mu}\,.
\ee
Separating the logarithm argument into a $t$-dependent and a $t$-independent 
part, one gets a logarithmic term not-involving time which gives
a \emph{conservative} 
contribution to the force in eq.~(\ref{eq:resreg}) which shifts 
logarithmically the mass of the binary system. The logarithmic
mass-shift $\delta M^{(R)}$
can be determined by requiring that its time derivative balance the acceleration
shift given by eq.~(\ref{eq:resreg}) \cite{Blanchet:2010zd} 
\be
\label{eq:mShift}
\frac{{\rm d}(\delta M^{(R)})}{{\rm d}t}=-\sum_A m_A\delta\ddot x_{Ai}\dot x_{Ai}
\,.
\ee
Substituting eq.~(\ref{eq:resreg}) into eq.~(\ref{eq:mShift}) and using the 
leading order quadrupole moment expression in eq.~(\ref{eq:Qlead}) allows
to turn the right hand side of eq.~(\ref{eq:mShift}) into a total time
derivative, enabling to identify the logarithmic mass shift as 
\cite{Blanchet:2010zd}
\be
\label{eq:dE}
\delta M^{(R)}|_{log}=
-\frac{2G_N^2M}5\pa{2Q_{ij}^{(5)}Q_{ij}^{(1)}-2Q_{ij}^{(4)}Q_{ij}^{(2)}+Q_{ij}^{(3)}Q_{ij}^{(3)}}
\log(\mu)\,.
\ee
Eq.~(\ref{eq:dE}) can be rewritten as a renormalization group flow equation
\cite{Goldberger:2012kf}
\be
\mu\frac{\rm d}{{\rm d}\mu}M^{(R)}(t,\mu)=-\frac{2G_N^2M}5\pa{2Q_{ij}^{(5)}Q_{ij}^{(1)}
-2Q_{ij}^{(4)}Q_{ij}^{(2)}+Q_{ij}^{(3)}Q_{ij}^{(3)}}\,.
\ee
This classical renormalization of the mass monopole term (which can be
identified with the Bondi mass of the system, that does not include the energy 
radiated to infinity) is explained in 
\cite{Goldberger:2012kf} by considering that the emitted radiation is 
scattered by the curved space and then absorbed, hence observers at different 
distance from the source would not agree on the value of the mass.

The ultraviolet nature of the divergence points to the incompleteness of the
effective theory in terms of multipole moments: the terms analytic in $\omega$
in eq.~(\ref{eq:radReacRes}) are sensitive to the short distance physics and
their actual value should be obtained by going to the theory at orbital radius.

The tail term radiation reaction force
is responsible for a conservative force at 4PN (as the leading radiation
reaction acts at 2.5PN and the tail term is a 1.5PN correction to it), so it
must be added to the conservative dynamics coming from the calculation
of the effective action not involving gravitational radiation, and indeed it
is responsible for the logarithmic term in eq.~(\ref{eq:Ec4PN}).

\subsection{Emitted flux}

\begin{figure}
  \begin{minipage}[htb]{.49\linewidth}
    \begin{center}
      \includegraphics[width=\linewidth]{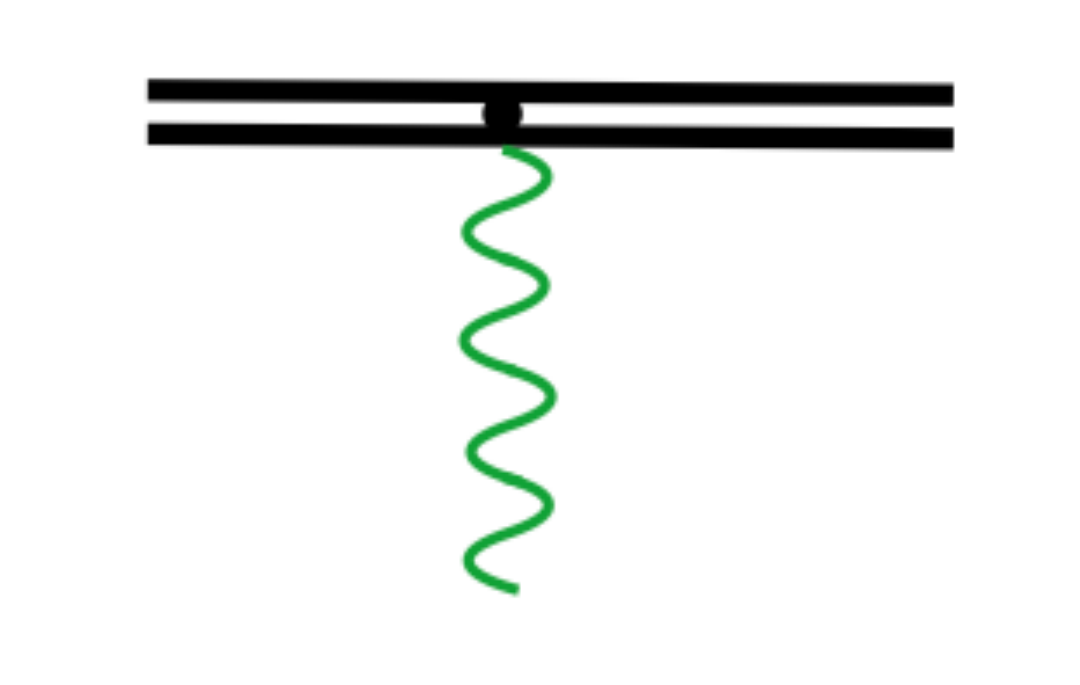}
    \end{center}
    \caption{Diagram representing the emission of a GW from a quadrupole source}
    \label{fi:quadGWemi}
  \end{minipage}
  \begin{minipage}{.49\linewidth}
    \begin{center}
      \includegraphics[width=\linewidth]{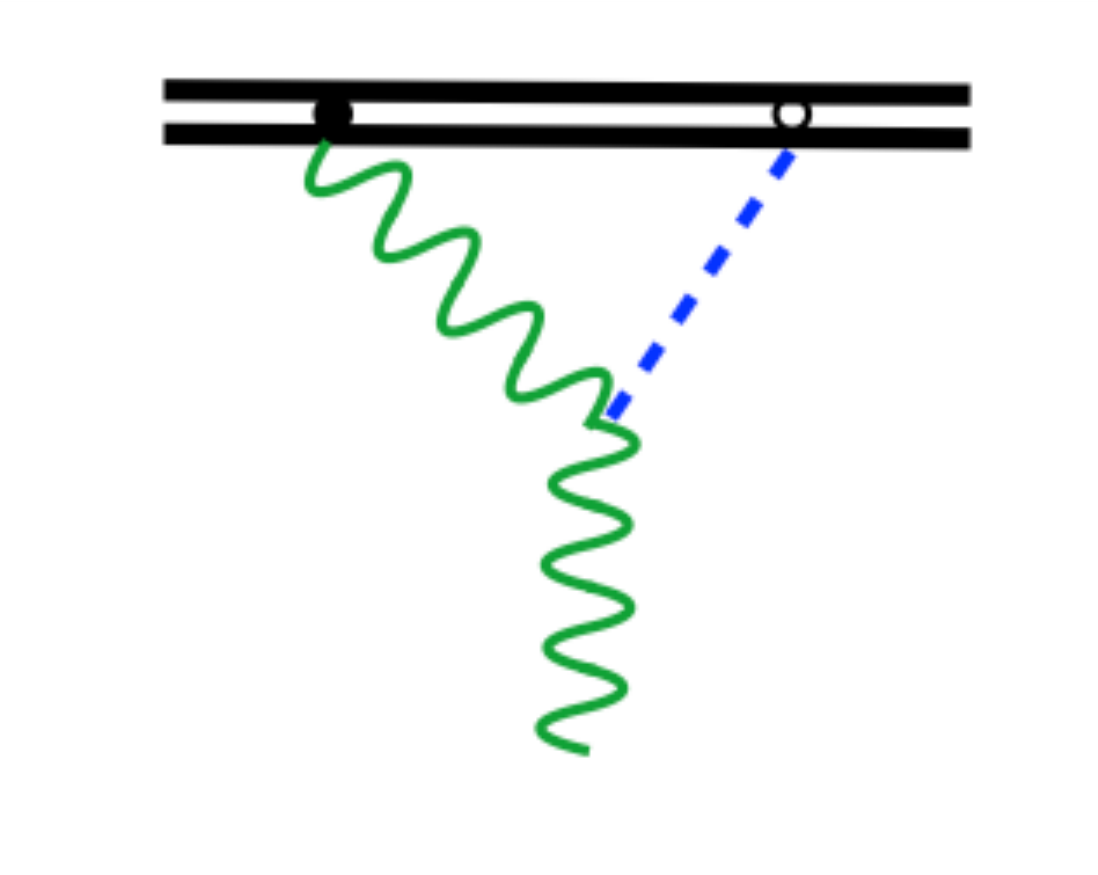}
      \caption{Emission of a GW from a quadrupole source with post-Minkowskian
      correction represented by the scattering off the background curved by
      the presence of binary system.}
      \label{fi:quadGWMemi}
    \end{center}
  \end{minipage}
\end{figure}

We have now shown how to perform the matching between the theory of extended
objects with multipoles and the theory at the orbital scale.
Taking the action for extended bodies in eq.~(\ref{eq:ext}) as a starting point,
the emitted GW-form and the total radiated power can be computed in terms of 
the source multipoles by evaluating the probability amplitude $A_h(\K)$ to emit 
a GW of 4-momentum $(\omega=|\K|,\K)$ and helicity $h$, using Feynman diagrams with 
one external radiating gravitational particle.
At leading order the amplitude for the emission of a GW with 3-momentum $\K$,
helicity $h$ and polarization 
tensor $\epsilon_{ij}$, is given by the diagram in 
fig.~\ref{fi:quadGWemi} and results in
\be
\label{eq:ampGW}
A_h(\K)=\frac{\K^2}{4\Lambda}Q_{ij}\epsilon^*_{ij}(\K,h)\,.
\ee

The GW-form can be computed using the closed time path formalism
\be
\!\!\!\!\!\!\! \sigma_{ij}(t,x)\supset-2G_N\Lambda_{ij;kl}\int {\rm d}t'{\rm d}^dx'\,G_R(t-t',x-x')\paq{\ddot I_{kl}+\frac 43\epsilon_{lmn}\dot J_{mn,k}-\frac 13\ddot I_{klm,m}}\,,\nonumber\\
\ee
where we have introduced the TT-projector $\Lambda_{ij;kl}$ defined as
\be
\Lambda_{ij;kl}=P_{ik}P_{jl}-\frac 1{d-1}P_{ij}P_{kl}\,,\qquad P_{ij}\equiv
\delta_{ij}-n_in_j\,,
\ee
being $n_i$ the unit vector in the direction of observation, and analog 
formulae hold for the following multipoles.
Analogously to what shown in the previous subsection, we have to take into
account the GW interaction with the
space time curvature produced by the source itself.
Including such effect give rise to a \emph{tail} effect, accounted by the 
diagram in fig.~\ref{fi:quadGWMemi}, which gives a contribution to the GW
amplitude and phase \cite{Blanchet:1993ec,Porto:2012as}
\be
\label{eq:htailI}
\sigma_{ij}\supset\Lambda_{ij;kl}\frac{2G}{r}\int \frac{{\rm d}\omega}{2\pi}
e^{i\omega(t-r)+iG_NM\omega\paq{\frac 1{\epsilon}+\log\pa{\frac\omega\mu}^2+\gamma-\frac{11}6}}
\pa{1+G_Nm|\omega|\pi}I_{kl}\,.
\ee
The infra-red singularity in the phase of the emitted wave
is un-physical as it can be absorbed in a re-definition of time in 
eq.~(\ref{eq:htailI}). 
Moreover any experiment, like LIGO and Virgo for 
instance, can only probe phase \emph{differences} (e.g. the GW 
phase difference between the instants when the wave enters and exits the 
experiment sensitive band) and the un-physical dependencies on the regulator 
$\epsilon$ and on the subtraction scale $\mu$ drops out of any observable.

The contribution from the magnetic quadrupole is analogous to the one in 
eq.~(\ref{eq:htailI}), and it is \cite{Porto:2012as}
\be
\label{eq:htailJ}
\!\!\!\!\!\!\!\!\!\!\!\!
\sigma_{ij}\supset\Lambda_{ij;kl}\frac{2G}{r}\int \frac{{\rm d}\omega}{2\pi}
e^{i\omega(t-r)+iG_NM\omega\paq{\frac 1\epsilon+\log\pa{\frac\omega\mu}^2+\gamma-\frac{7}3}}
\pa{1+G_Nm|\omega|\pi}J_{kl}
\ee
where the finite number associated with the logarithm is still un-physical,
as it depends on the choice of the arbitrary scale $\mu$, but 
the difference between the terms in the phase in eqs.~(\ref{eq:htailI},
\ref{eq:htailJ}) \emph{is} physical, as $\mu$ can be chosen only once
\cite{Porto:2012as}.
Spin effects can be included straightforwardly by using the appropriate 
multipole expression.

\begin{figure}
  \begin{center}
    \includegraphics[width=.45\linewidth]{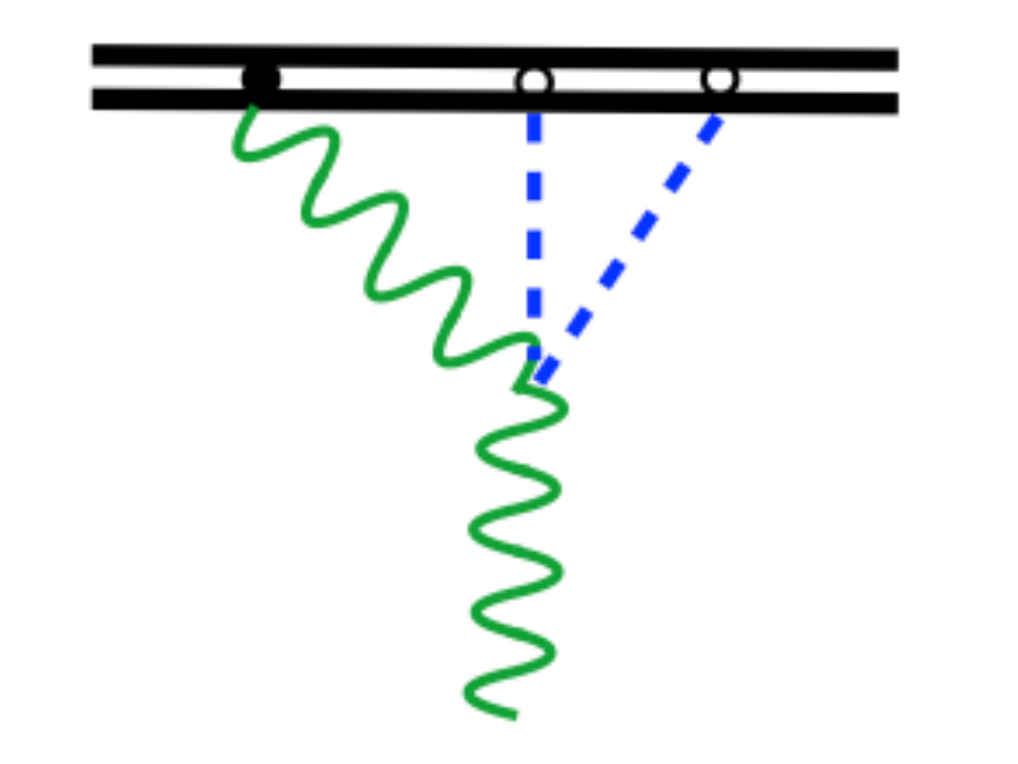}
    \includegraphics[width=.45\linewidth]{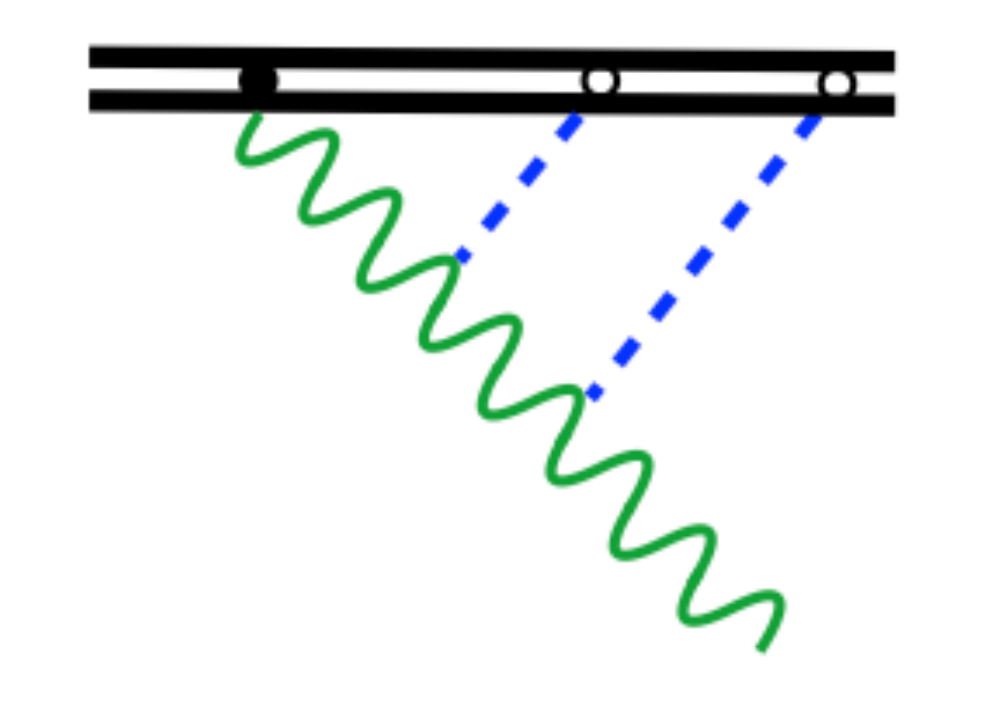}
    \caption{Diagrams contributing to $A_h(k)|_{v^6}$.}
    \label{fi:GWMM}
  \end{center}
\end{figure}

The total emitted flux can be computed once the amplitude of the GW has been
evaluated, via the standard formula
\be
P=\frac{r^2}{32\pi G_N}\int {\rm d}\Omega\,\langle\dot h_{ij}\dot h_{ij}\rangle\,,
\ee
but there is actually a shortcut, as the emission energy rate can be computed
directly from the amplitude $A_h(\K)$ without solving for $\sigma_{ij}$
via the optical theorem formula
\be
\label{eq:optical}
{\rm d}P_h(\omega)=\frac 1T\frac{{\rm d}^3{\bf k}}{(2\pi)^3}|A_h(\K)|^2\,.
\ee
Using eqs.~(\ref{eq:ampGW}), (\ref{eq:optical}) and summing over polarizations one gets
\cite{Goldberger:2009qd}
\be
P\simeq\frac{G_N}{5\pi T}\int_0^\infty {\rm d}\omega \omega^6 \paq{|I_{ij}(\omega)|^2+
\frac{16}9|J_{ij}(\omega)|^2+\frac 5{189}k^2|I_{ijk}(\omega)|^2+\ldots}
\ee
which, once averaged over time, recovers at the lowest order the standard 
Einstein quadrupole formula $P=G_N\langle \dddot I_{ij}^2\rangle/5$.
There are however corrections to this result for any given multipole, due to 
the scattering of the GW off the curved space-time because of the presence of 
the static potential due to the presence of the massive binary system.
The first of such corrections scale as
$G_NM\int ({\rm d}^4k\frac 1k^2)^2\delta^3(k)\sim G_NMk\sim v^3$ (for radiation
$k\sim v/r$), that is a 1.5PN correction with respect to the leading order.
The tail amplitude is described by the diagram in fig.~\ref{fi:quadGWMemi} and 
it adds up to the leading order to give a contribution to the flux going as
\be
\left|\frac{A_h|_{v^3}}{A_h|_{v^0}}\right|^2=1+2\pi G_NM \omega+O(v^6)\,.
\ee
The diagrams quadratic in the background curvature are portraited in 
figs.~\ref{fi:GWMM} and they give an ultraviolet divergence, with a logarithmic
term \cite{Goldberger:2009qd}
\be
\left|\frac{A_h|_{v^6}}{A_h|_{v^0}}\right|_{v^6}^2=-(G_NM\omega)^2\frac{214}{105}
\ln\frac{\omega^2}{\mu^2}+\ldots\,,
\ee
depending on the arbitrary subtraction scale $\mu$, where finite contributions
have been omitted. This short-distance 
singularity represents a failure of the effective theory at the radiation scale
to correctly describe short-distance physics: in order to fix the omitted
numerical quantity analytic in $\omega$ one should match the multipole theory 
to the theory in which the binary constituents are at a finite distance $r$.

However the coefficient of the logarithm is physical and we can then proceed
to renormalize the theory at the radiation scale, which
is done in the usual fashion as in quantum field theory, although here the
effect is completely classical.
Since $|A_h|^2$ enters physical results like energy emission, it should be
independent of the arbitrary scale $\mu$: this can only happen if we assume
a $\mu$ dependence on the \emph{renormalized} multipole moments $I_{ij}$ of the
type:
\be
\label{eq:renI}
\mu\frac{{\rm d}I^{(R)}_{ij}}{{\rm d}\mu}=-\frac{214}{105}(G_NM\omega)^2I_{ij}^{(R)}\,.
\ee
Assuming that $A_{h}$ is expressed in terms of the $I^{(R)}_{ij}$, the total
dependence of $|A_h|^2$ on $\mu$ cancels out (it makes no difference if using
$I^{(R)}_{ij}$ or the ``bare" $I_{ij}$ in $A_h|_{v^3}$, as the difference is
higher order in $v$).
The background curvature has the effect of ``smearing'' the multipole source
which cannot be considered perfectly localized at the origin of the coordinates:
the value of the $I^{(R)}_{ij}$ will depend on the scale at which the observer
will measure it.

A consequence of this result is that eq.~(\ref{eq:renI}) admits a solution
\be
\label{eq:RGsol}
I^{(R)}_{ij}(k,\mu)=\pa{\frac\mu{\mu_0}}^{-\frac{214}{105}(G_NM\omega)^2}I_{ij}(k,\mu_0)
\ee
that constrains the patterns of logarithms that can appear at higher orders.
Once the multipole is known at some scale, like the orbital scale separation, 
then it can be known at any other scale by virtue of eq.~(\ref{eq:RGsol}).

\begin{figure}
  \begin{center}
    \includegraphics[width=.6\linewidth]{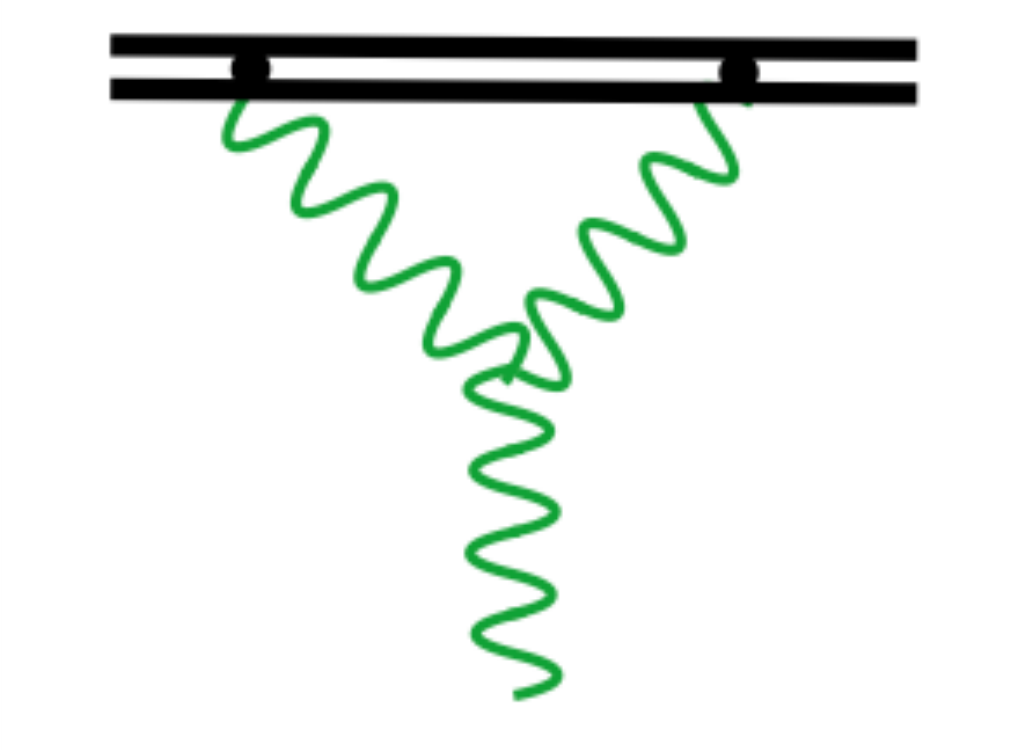}
  \end{center}
  \caption{Memory diagram: GW emitted from a source scattered by another
  GW before reaching the observer.}
  \label{fi:memory}
\end{figure}

Finally one could consider the scattering of the emitted GW wave off another
GW, as in fig.~\ref{fi:memory}. This process is known as
\emph{non-linear memory} effect,
it represents a 2.5PN correction with respect to the leading emission
amplitude \cite{Christodoulou:1991cr,Blanchet:1992br,Blanchet:1997jj}
and it has not yet been computed within the effective field theory formalism.

Combined tail and memory effects enter at 4PN order in the emitted radiation,
i.e. double scattering of the emitted radiation off the background curvature
\emph{and} off another GW. The divergences describing such process have been
analyzed in \cite{Goldberger:2012kf}, leading to the original derivation of the
mass renormalization described in subsec.~\ref{sse:radreac}.
The renormalization group equations allow a resummation of the logarithmic term
making a non-trivial prediction for the pattern of the leading UV logarithms
appearing at higher orders \cite{Goldberger:2009qd,Goldberger:2012kf}.

\section{Conclusions}

This Topical Review aims at giving an overview of the basic ideas
of Effective Field Theory methods proposed in \cite{Goldberger:2004jt} to model
gravitationally bound, inspiralling compact binary systems.
The study of such system has both phenomenological and theoretical motivations,
due to the forthcoming observational campaign of the large interferometric detectors
LIGO and Virgo (and eventually KAGRA and Indigo) on one side, and on the development
of efficient numerical methods to solve Einstein equations on the other side.

The post-Newtonian investigation of the compact binary inspiral problem has a long history
in analytical perturbative solutions of the Einstein equations, but EFT methods
have allowed a new field theory insight into it.
The problem admits a description in terms of well separated scales
(the individual source size, the binary component distance and the radiation 
wavelength),
with a single dimension-less perturbative parameter (at least in the binary black-hole case),
represented by the relative velocity of the individual components of the binary
system.
The EFT methods allow to treat in a single, powerful framework both conservative
and dissipative
effects and provide efficient tools to compute observable quantities.
They give an organizational principle for performing a systematic expansion in
the PN perturbative parameter. The scale factorization is already evident at 
the level of the action,
which allows a considerable computational simplification with respect to 
methods working at the level of the equations of motion.
The effective field theory approach reviewed here has much in common with standard
quantum field theory techniques because of the common underlying field theory 
structure and it is completely classic.

Physics at different scales are related by renormalization group flow, and
all kind of divergences, arising from incomplete knowledge of the underlying 
short-distance physics as well as from long-distance effects and from gauge 
artifacts, are technically treated
on equal footing via dimensional regularization.
Indeed the use of field theory since several decades has allowed the development
of powerful tools to address all the technical problems (like handling of 
divergences and computation of Feynman integrals) on the computational side.

Finally, the existence of an additional independent method to compute physical
observables of the binary problem in General Relativity is welcome \emph{per se},
as it allows an independent check of computations of formidable complexity.

\ack

SF is supported by the Fonds National Suisse, RS is supported by the FAPESP
grant 2013/04538-5. 
RS wishes to thank the CERN theory division for hospitality and support during
the last stage of this work.

\section*{References}

\bibliography{CQGreviewFoffaSturani}

\end{document}